\documentclass[preprint,authoryear]{elsarticle}
\usepackage{amsmath, amssymb, amsfonts}
\usepackage{graphicx}
\usepackage{subfigure}
\usepackage{color}
\usepackage[ruled]{algorithm}
\usepackage{algpseudocode}
\usepackage{rotating}
\usepackage[table]{xcolor}
\usepackage{pdflscape}
\usepackage{multirow}
\usepackage{hyperref}
\usepackage{adjustbox}
\usepackage{caption}
\usepackage{booktabs}
\usepackage{fixltx2e}
\usepackage{natbib}
\journal{Elselvier}

\begin{document}

% new commands
\newcommand{\rmnum}[1]{\romannumeral #1}
\newcommand{\presence}[1]{\ensuremath{{\texttt{PresenceOf}}\!\!\left(#1\right)}}
\newcommand{\cumulative}[3]{\ensuremath{{\texttt{Cumulative}}\!\left(#1,#2,#3\right)}}
\newcommand{\overbar}[1]{\mkern 1.5mu\overline{\mkern-1.5mu#1\mkern-1.5mu}\mkern 1.5mu}
\newcommand{\spaceX}{\!\!\!\!\!\!}
\newcommand{\spaceS}{\!\!}

% lower bounds
\newcommand{\DMV}{$\underline{LB}$}
\newcommand{\DMVextra}[1]{$\underline{LB}\!\left(#1\right)$}
\newcommand{\BSP}{\textsl{LB\ensuremath{\textsubscript{1}}}}
\newcommand{\PH}{\textsl{LB\ensuremath{\textsubscript{3}}}}
\newcommand{\LBRELAX}{\textsl{LB\ensuremath{\textsubscript{2}}}}
\newcommand{\RMIP}{\textsl{RELAX}}

% upper bounds
\newcommand{\MIP}{\textsl{EXACT}}
\newcommand{\ASSIGN}{\textsl{ASSIGN}}
\newcommand{\FG}{\textsl{FF}}
\newcommand{\Alga}{\textsl{APPROX}}
\newcommand{\CPr}{\textsl{PACK}}
\newcommand{\APr}{\textsl{HEUR}}
\newcommand{\APrr}{\textsl{HEUR}\ensuremath{'}}
\newcommand{\U}{\textsl{UB}}
\newcommand{\UbFunction}[1]{\ensuremath{b}(#1)}
\newcommand{\GA}{\textsl{MXGA}}

% parameters
\newcommand{\TPACK}{$t_{lim}^{\text{\textsl{PACK}}}$}
\newcommand{\LHEUR}{$a_{lim}^{\text{\textsl{HEUR}}}$}
\newcommand{\LHEURp}{$a_{lim}^{\text{\textsl{HEUR'}}}$}

% variables
\newcommand{\thetav}[1]{\theta^{\text{\rmnum{#1}}}_{ee'}}

% other notation
\newcommand{\BP}{BP}
\newcommand{\BPPDD}{2BPP with DD}
\newcommand{\co}{\textsl{count}}
\newcommand{\RLS}{\textsl{RLS}}
\newcommand{\MIPG}{MIP}
\newcommand{\CP}{CP}

\begin{frontmatter}

\title{A Hybrid Feasibility Constraints-Guided Search to the Two-Dimensional Bin Packing Problem with Due Dates}
\author[a1]{Sergey Polyakovskiy} \ead{Sergey.Polyakovskiy@adelaide.edu.au}
\author[a2]{Rym M'Hallah\corref{cor}} \ead{rymmha@yahoo.com}
\cortext[cor]{Corresponding author. Tel: +965-669-14150. Fax: +965-248-37332.}
\address[a1]{Optimisation and Logistics Group, School of Computer Science, University of Adelaide, Australia.}
\address[a2]{Department of Statistics and Operations Research, College of Science, Kuwait University, P.O. Box 5969, Safat 13060, Kuwait.}

\begin{abstract}
The two-dimensional non-oriented bin packing problem with due dates packs a set of rectangular items, which may be rotated by 90 degrees, into identical rectangular bins. The bins have equal processing times. An item's lateness is the difference between its due date and the completion time of its bin. The problem packs all items without overlap as to minimize maximum lateness $L_{max}$.

The paper proposes a tight lower bound that enhances an existing bound on $L_{max}$ by 31.30\% for 24.07\% of the benchmark instances and matches it in 30.87\% cases. Moreover, it models the problem via mixed integer programming (MIP), and solves small-sized instances exactly using \textsl{CPLEX}. It approximately solves larger-sized instances using a two-stage heuristic. The first stage constructs an initial solution via a first-fit heuristic that applies an iterative \emph{constraint programming} (CP)-based neighborhood search. The second stage, which is iterative too, approximately solves a series of assignment low-level MIPs that are guided by \emph{feasibility constraints}. It then enhances the solution via a high-level random local search. The approximate approach improves existing upper bounds by 27.45\% on average, and obtains the optimum for 33.93\% of the instances.  Overall, the exact and approximate approaches find the optimum in 39.07\% cases.

The proposed approach is applicable to complex problems. It applies CP and MIP sequentially, while exploring their advantages, and hybridizes heuristic search with MIP. It embeds a new lookahead strategy that guards against infeasible search directions and constrains the search to improving directions only; thus, differs from traditional lookahead beam searches.
\end{abstract}


\end{frontmatter}

%%%%%%%%%%%%%%%%%%%%%%%%%%%%%%%%%%%%%%%%%%%%%%%%%%%%%%%%%%%%%%%%%%%%%%%%%%%%%%%%%%%%%%%%%%%%%%%%%%%%%%%%%%%%%%%%%%
\section{Introduction}
%%%%%%%%%%%%%%%%%%%%%%%%%%%%%%%%%%%%%%%%%%%%%%%%%%%%%%%%%%%%%%%%%%%%%%%%%%%%%%%%%%%%%%%%%%%%%%%%%%%%%%%%%%%%%%%%%%
Bin packing ({\BP}) is a classical strongly $\mathcal{NP}$-hard combinatorial optimization problem \citep{Jansen,Johnson}. It consists in packing a set of items into as few bins as possible. Because of its prevalence in industry, {\BP} has engendered many variants. Some variants impose additional constraints on the packing of the items or on the types of bins such as the oriented, orthogonal, guillotine, and variable-sized {\BP}. More recent variants combine {\BP} with further complicating combinatorial aspects. For example, {\BP} appears in combination with routing problems: minimizing transportation costs subject to loading constraints \citep{Iori13}. It also emerges in lock scheduling ~\citep{Verstichel15} where lockages are scheduled, chambers are assigned to ships, and ships are positioned into chambers.

Following this trend, this paper addresses the non-oriented two-di\-men\-sio\-nal {\BP} problem where items have due dates. This problem, denoted hereafter {\BPPDD}, searches for a feasible packing of a given set of $n$ rectangular items into a set of at most $b \leq n$ identical rectangular bins, and schedules their packing as to minimize the maximum lateness $L_{max}$ of the items. Each item is characterized by its width, height, and due date. Its lateness is the difference between its completion time and its due date, where its completion time is that of its assigned bin. All bins' processing times are equal regardless of their assigned items. This problem is common in make-to-order low-volume production systems such as the high-fashion apparel industry and food delivery. In these contexts, packing efficiency might be increased by mixing up several orders; however, the increased efficiency can not be at the cost of customer service. That is, a company should choose, from the pool of items emanating from all orders, the ones that need to be packed (or cut) simultaneously with the objective of maximizing material utilization (or packing efficiency) while meeting due dates.

Similar problems were considered in the literature. \citet{Reinertsen10} investigate the one-dimensional cutting stock problem within steel manufacturing where orders have due dates that must be met. \citet{Arbib2017} study a one-dimensional bin packing problem with the objective of minimizing a weighted sum of maximum lateness and maximum completion time. \citet{Li96} tackles a two-dimensional cutting stock problem where meeting the orders' due dates is more important than minimizing the wasted material. \citet{Arbib14} survey the state of the art on packing with due dates. \citet{Polyakovskiy11} address the on-line cutting of small rectangular items out of large rectangular stock material using parallel machines in a just-in-time environment. The cutting pattern minimizes both material waste and the sum of earliness-tardiness of the items. \citet{Polyakovskiy:2017} consider another variant of {\BP}. Items that are cut from the same bin form a batch, whose processing time depends on its assigned items. The items of a batch share the completion time of their bin. The problem searches for the cutting plan that minimizes the weighted sum of the earliness and tardiness of the items.

\citet{Bennell13} deal with a bi-criteria version of {\BPPDD}. They minimize simultaneously the number of used bins and the maximum lateness of the items. They propose a lower bound {\BSP} to $L_{max}$ and approximately solve their bi-criteria problem using a single-crossover genetic algorithm, a multi-crossover genetic algorithm ({\GA}), a unified tabu search, and a randomized descent. They generate a benchmark set for which they report the best average value for each of their two objective functions. They conclude that {\GA} yields consistently the best upper bound $L_{max}^{\GA}$ on $L_{max}$.

This paper focuses on minimizing $L_{max}$ only (in lieu of both the number of used bins and $L_{max}$ as the bi-criteria case does). It is in no way a shortcoming for four reasons. First, the number of bins is naturally bounded. Second, for a given feasible bound on the number of used bins $b$, searching for the minimal $L_{max}$ is a standard practice to tackle multi-objective problems. Thus, it can be applied iteratively to build the Pareto optimal frontier of the bi-criteria problem. Third, the objectives of minimizing lateness and maximizing packing efficiency do not necessarily conflict \citep{Bennell13}. Finally, it can be used by decision makers as a decision support tool that quantifies the tradeoff between service quality loss and reduction of both ecological cost and waste material.

As for all difficult combinatorial optimization problems, finding an exact solution, in a reasonable time, for large-sized instances of {\BPPDD} is challenging. Indeed, {\BP} variants are generally tackled using approximate approaches that are based on meta-heuristics \citep{Lodi2002379, Lodi2014107, Sim20131549}, including genetic algorithms, and hyper-heuristics \citep{Burke,Lopez,Sim201537}. Unlike the aforementioned techniques, the proposed two-stage approximate approach for {\BPPDD} explores the complementary strengths of constraint programming ({\CP}) and mixed integer programming ({\MIPG}). In its first stage, it applies {\CP}. In its second stage, it hybridizes heuristic search with {\MIPG}, where {\MIPG} is in turn guided by feasibility constraints. In addition, it applies an innovative lookahead strategy that (i) forbids searching in directions that will eventually lead to infeasible solutions and (ii) directs the search towards improving solutions only. Consequently, the proposed assignment based packing approach with its new lookahead strategy is a viable alternative to the constructive heuristics traditionally applied to {\BP}, where bins are filled sequentially in a very greedy manner \citep{Lodi2002379}.

Section \ref{sec:prob} gives a mathematical formulation of the {\BPPDD}. Section \ref{sec:Background} provides essential background information on feasibility constraints and on a {\CP}-based approach for the two-dimensional orthogonal packing problem. Section \ref{sec:LB} presents the existing lower bounds {\BSP} and {\LBRELAX} and the new one {\PH}. Section \ref{sec:UB} proposes the two-stage solution approach with Section \ref{sec:FFC} detailing the first-fit heuristic (i.e., stage one), Section \ref{sec:AP} describing the assignment based heuristic, and Section \ref{sec:SC} summarizing the second stage. Section \ref{sec:CE} discusses the results of the computational investigation performed on benchmark instances. Finally, Section \ref{sec:Conclusion} summarizes the paper and gives some concluding remarks.

%%%%%%%%%%%%%%%%%%%%%%%%%%%%%%%%%%%%%%%%%%%%%%%%%%%%%%%%%%%%%%%%%%%%%%%%%%%%%%%%%%%%%%%%%%%%%%%%%%%%%%%%%%%%%%%%%%
\section{Mathematical Formulation}\label{sec:prob}
%%%%%%%%%%%%%%%%%%%%%%%%%%%%%%%%%%%%%%%%%%%%%%%%%%%%%%%%%%%%%%%%%%%%%%%%%%%%%%%%%%%%%%%%%%%%%%%%%%%%%%%%%%%%%%%%%%
Let $B=\left\{1,\ldots,b\right\}$ be a set of $b$ identical rectangular bins. Bin $k \in B$ has a width $W$, a height $H$, and a processing time $P$. Items assigned to the same bin have a common completion time. Let $N=\left\{1,\ldots,n\right\}$ be a set of $n$ rectangular items, where $n \geq b$. Item $i \in N$ has a width $w_i \leq W$, a height $h_i \leq H$, and a due date $d_i$. When $w_i \leq H$ and $h_i \leq W$, item $i \in N$ may be rotated by 90\textsuperscript{o} for packing purposes. Every item $i \in N$ must be packed without overlap and must be completely contained within its assigned bin.  When assigned to bin $k$, item $i \in N$ has a completion time $C_i=kP$ and lateness $L_i=C_i-d_i$. The {\BPPDD} consists in finding a feasible packing of the $n$ items into the available bins with the objective of minimizing $L_{max},$ defined by $L_{max}=\displaystyle\max_{i \in N} \left\{C_i-d_i\right\}$.

Let $N^*$ denote the set $N$ appended by the rotated duplicates. The duplicate of item $i,\ i \in N$, is item $n+i$ of width $h_i,$ height $w_i$ and due date $d_i$. The problem is then modeled as an {\MIPG} with six types of variables.
\begin{itemize}
\setlength\itemsep{0em}
\item $\textbf{\textit{x}}$ and $\textbf{\textit{y}}$ denote the position of an item within its assigned bin, where $x_i\geq 0$ and $y_i\geq 0,\ i \in N^*,$ are the bottom left coordinates of item $i$.
\item $\textbf{\textit{f}}$ signals the assignment of an item to a bin, where $f_{ik}=1$ if item $i$ is packed into bin $k,\ i\in N^*,\ k\in B$, and 0 otherwise.
\item $\textbf{\textit{l}}$ and $\textbf{\textit{u}}$ are binary. They refer to the relative position of two items. $l_{ij}=1$ (resp. $u_{ij}=1$), $i\in N^*,\ j \in N^*,\ i \neq j,\ j \neq i+n$, and $i \neq j+n$, is used to make $i$ to the left of (resp. below) $j$ when $i$ and $j$ are in the same bin.
\item The sixth is the objective value, which is $L_{max}$.
\end{itemize}
When the rotated duplicate of item $i$ cannot fit into a bin, i.e. its $w_i \leq H$ and $h_i \leq W,\ i \in N$, its corresponding decision variables are not defined; thus, they are omitted from the model; so are any corresponding constraints.

The {\MIPG} model ({\MIP}), which uses the disjunctive constraint technique of~\citet{Chen1995} and~\citet{Onodera1991}, follows.
{\footnotesize
\begin{flalign}
\mbox{min} \ & L_{max}& \label{eq:1}\\[-0.05cm]
\mbox{s.t.} \ & l_{ij} + l_{ji} + u_{ij} + u_{ji} - f_{ik} - f_{jk} \geq -1&(i,j) \in {N^*}^2,\ i < j, \ j \neq n+i,\ k\in B \label{eq:2}\\[-0.05cm]
& x_i+w_i \leq x_j + W\left(1-l_{ij}\right)& (i,j) \in {N^*}^2, \ i \neq j, \ j \neq i+n, \ i \neq j+n \label{eq:3}\\[-0.05cm]
& y_i+h_i \leq y_j + H\left(1-u_{ij}\right)& (i,j) \in {N^*}^2, \ i \neq j, \ j \neq i+n, \ i \neq j+n \label{eq:4}\\[-0.05cm]
& x_i \leq W - w_i& i \in N^* \label{eq:5}\\[-0.05cm]
& y_i \leq H - h_i& i \in N^* \label{eq:6}\\[-0.05cm]
& \sum_{k \in B} \left(f_{ik}+f_{n+i k}\right) = 1& i \in N\label{eq:7}\\[-0.05cm]
& \sum_{k \in B} \left(kP-d_i\right) f_{ik} \leq L_{max}& i \in N^* \label{eq:8}\\[-0.05cm]
& l_{ij}\in \left\{0,1\right\}, \; b_{ij} \in \left\{0,1\right\}& (i,j) \in {N^*}^2, \ i \neq j, \ j\neq i+n, \ i\neq j+n \label{eq:9}\\[-0.05cm]
& f_{ik}\in \left\{0,1\right\}& i \in N^*,\ k \in B \label{eq:10}\\[-0.05cm]
& x_i \in \mathbb{R}_{\geq0}, \; y_i \in \mathbb{R}_{\geq0}& i \in N^* \label{eq:11}\\[-0.05cm]
& L_{max} \in \mathbb{R} \label{eq:12}
\end{flalign}
}
Equation (\ref{eq:1}) defines the objective value. It minimizes the maximum lateness. Equation (\ref{eq:2}) determines the relative position of any pair of items that are assigned to a same bin: one of them is either left of and/or below the other. Equation (\ref{eq:3}) ensures that items $i$ and $j$ do not overlap horizontally if in the same bin while Equation (\ref{eq:4}) guarantees that they do not overlap vertically. Equations (\ref{eq:5}) and (\ref{eq:6}) guarantee that $i$ is entirely contained within a bin. Equation (\ref{eq:7}) ensures either $i$ or its rotated copy $i+n$ is packed into exactly one bin. Equation (\ref{eq:8}) sets $L_{max}$ larger than or equal to the lateness $L_i$ of $i$, where $L_i$ is the difference between the completion time of the bin to which $i$ is assigned and the due date of $i$. Finally, Equations (\ref{eq:9})-(\ref{eq:12}) declare the variables' types. The model has a quadratic number of variables in $n$. Because $b$ is bounded by $n,$ the model has a cubic number of constraints in $n$. The solution space has a large number of alternative solutions with many symmetric packing set ups. Subsequently, {\MIP} is hard to solve. Small-sized instances with as few as 20 items require significant computational effort.

%%%%%%%%%%%%%%%%%%%%%%%%%%%%%%%%%%%%%%%%%%%%%%%%%%%%%%%%%%%%%%%%%%%%%%%%%%%%%%%%%%%%%%%%%%%%%%%%%%%%%%%%%%%%%%%%%%
\section{Background}\label{sec:Background}
%%%%%%%%%%%%%%%%%%%%%%%%%%%%%%%%%%%%%%%%%%%%%%%%%%%%%%%%%%%%%%%%%%%%%%%%%%%%%%%%%%%%%%%%%%%%%%%%%%%%%%%%%%%%%%%%%%
The two-dimensional orthogonal packing problem (2OPP) determines whether a set of rectangular items can be packed into a rectangular bin. This decision problem is used, in this paper, when generating the lower bound {\PH} (cf. Section \ref{LB3}) and as part of the new first-fit heuristic ({\FG}) (cf. Section \ref{sec:FFC}) when searching for a feasible packing.

{\PH} is the optimal solution of a mixed integer program that substitutes the containment and overlap constraints of {\MIP} by feasibility constraints. These constraints explore the notion of dual feasible functions (DFF) to strengthen the resulting relaxation of {\MIP}. Section \ref{sec:fc} presents DFFs and explains their application to the non-oriented version of 2OPP.

2OPP arises also as a part of the constructive heuristic {\FG}, which constitutes the first stage of the proposed solution approach {\Alga}. Specifically, every time it considers a subset of items, {\FG} solves a non-oriented 2OPP to determine the feasibility of packing those items into a bin. As it calls the 2OPP decision problem several times, {\FG} needs an effective way of tackling it. For this purpose, it models the problem as a CP, and augments it with two additional constraints issued from two related non-preemptive cumulative scheduling problems. Section \ref{sec:CPBA} presents this CP model.

%>>>>>>>>>>>>>>>>>>>>>>>>>>>>>>>>>>>>>>>>>>>>>>>>>>>>>>>>>>>>>>>>>>>>>>>>>>>>>>>>>>>>>>>>>>>>>>>>>>>>>>>>>>>>>>>>>
\subsection{Feasibility Constraints} \label{sec:fc}
\citet{Alves2016:2006} explore standard DFFs for different combinatorial optimization problems, including cutting and packing problems. \citet{Fekete04} apply DFFs to find a lower bound $L_{2d}$ to the minimal number of bins needed to pack orthogonally a given set of two-dimensional oriented items. This section explains how to use DFFs to generate feasibility constraints.

A function $u: \left[0,1\right] \rightarrow \left[0,1\right]$ is dual feasible if
$$\displaystyle\sum_{s \in S} s \leq 1 \Rightarrow \displaystyle\sum_{s \in S} u\left(s\right) \leq 1$$
holds for any set $S$ of non-negative real numbers. Let $u_1$ and $u_2$ be two valid DFFs. For the problem at hand, the DFFs transform the scaled sizes $(w_i',h_i')$ of item $i \in I$ into differently scaled ones $(u_1(w_i'),u_2(h_i')) \in \left(0,1\right]$ where $w'_i = w_i/W$ and $h'_i = h_i/H.$ For a feasible packing into a single bin to exist, the sum of the areas of the transformed items must be less than or equal to 1,
\begin{eqnarray}
&\displaystyle\sum_{i \in I} u_1 \left(w'_i\right) u_2 \left(h'_i\right) \leq 1. \label{dff:F}
\end{eqnarray}
This section explains how DFFs are combined in various ways to generate $m$ inequalities/constraints in the form of Equation (\ref{dff:F}) for the non-oriented 2OPP.

Let $\mathcal{A}^o=(\alpha^o_{ci}) \in {R}_{\geq 0}^{m \times n}$ and $\mathcal{A}^r=(\alpha^r_{ci}) \in {R}_{\geq 0}^{m \times n}$ denote two real-valued technological matrices. Element $\alpha^o_{ci}$ (resp. $\alpha^r_{ci}$), $i \in N,\ c=1,\ldots,m$, is a scaled area computed using $w'_i$ and $h'_i$ (resp. $w"_i=h_i/W$ and $h"_i=w_i/H$) as arguments for DFFs $u_1$ and $u_2$, respectively. \citet{Fekete04} designed DFFs, namely $u^{\left(1\right)}$, $U^{\left(\epsilon\right)}$, and $\phi^{\left(\epsilon\right)}$, $\epsilon = p,q$. The functions and approach they use to obtained $L_{2d}$ is used herein to derive the combinations of functions $(u_1,u_2)$. The DFFs' input parameters $(p,q) \in \left(0,0.5\right]^2,$ as further specified in Section \ref{sec:CE}.

Furthermore, let $\textbf{t}^o \in \left\{0,1\right\}^n$ (resp. $\textbf{t}^r \in \left\{0,1\right\}^n$) be a binary decision vector such that $t^o_i = 1$ (resp. $t^r_i = 1$) if item $i,\ i \in N$, is packed into the bin without rotation (resp. with rotation) and 0 otherwise. When $t^o_i + t^r_i \leq 1,$ the inequality
\begin{eqnarray}
&\displaystyle\sum_{i \in N} \left(\alpha^o_{ci} t^o_i + \alpha^r_{ci} t^r_i \right)\leq 1,& c=1,\ldots,m, \label{dff:FC}
\end{eqnarray}
derived from Equation (\ref{dff:F}), is a valid feasibility constraint.
Equation (\ref{dff:FC}) assumes that the rotated duplicate of an item $i,\ i \in N,$ can fit into the bin. As mentioned in Section \ref{sec:prob}, when this assumption does not hold, $t^r_i=0$ and is omitted from Equation (\ref{dff:FC}).

Some of the $m$ constraints of Equation (\ref{dff:FC}) may be redundant.  A constraint $c,\ c = 1,\ldots,m,$ is redundant if either $\displaystyle\sum_{i \in N} \max{\left(\alpha^o_{ci},\alpha^r_{ci}\right)}\leq 1$ or there exists $c',\ c'=1,\ldots,m,\ c \neq c',$ such that both $\alpha^o_{ci} \leq \alpha^o_{c'i}$ and $\alpha^r_{ci} \leq \alpha^r_{c'i}$ for all $i \in N$.

%>>>>>>>>>>>>>>>>>>>>>>>>>>>>>>>>>>>>>>>>>>>>>>>>>>>>>>>>>>>>>>>>>>>>>>>>>>>>>>>>>>>>>>>>>>>>>>>>>>>>>>>>>>>>>>>>>
\subsection{Solving the 2OPP with Constraint Programming}\label{sec:CPBA}

This section develops a CP model for the non-oriented 2OPP.
The CP model, which is an extension of the model of \citet{Clautiaux08} for orthogonal packing, is strengthened by constraints issued of two non-preemptive cumulative scheduling problems. In this model, a bin corresponds to two distinct resources $r_w$ and $r_h$ of capacity $W$ and $H$, respectively, while the items to two sets of activities $A^w=\left\{a_1^w,\ldots,a_{2n}^w\right\}$ and $A^h=\left\{a_1^h,\ldots,a_{2n}^h\right\}$ where $a_i^w$ and $a_i^h$ are the width and height of item $i,\, i \in N^*$. The first (resp. second) scheduling problem treats the widths (heights) of the items as processing times of activities $A^w$ (resp. $A^h$) and considers the heights (widths) of the items as the amount of resource $r_h$ ($r_w$) required to complete these activities. The activities in $A^w$ and $A^h$ have compatibility restrictions; i.e., $a_i^{\bullet}$ and $a_{i+n}^{\bullet},\ \bullet=w,\, h,$ cannot both be scheduled.

The first (resp. second) scheduling problem investigates whether its set of activities $A^w$ (resp. $A^h$) can be performed within their respective time windows, without preemption and without exceeding the availability $H$ (resp. $W$) of required resource $r_h$ (resp. $r_w$). In fact, $A^w$ and $A^h$ are to be performed concurrently but using two different resources. Activity $a_i^w$ has a processing time $w_i$ and a time window $\left[0,W-w_i\right)$. To be processed, it uses an amount $h_i$ of resource $r_h$. Similarly, activity $a_i^h$ has a processing time $h_i$ and a time window $\left[0,H-h_i\right)$. Its processing requires an amount $w_i$ of resource $r_w$. Let $s_i^w$ and $s_i^h$ denote the respective starting times of activities $a_i^w$ and $a_i^h$. Then, $s_i^w$ and $s_i^h$ are the coordinates $(x_i,y_i)$ of item $i$ in the bin. The CP model that solves 2OPP is then given as:

{\footnotesize
\begin{flalign}
&\!
\left( \presence{a_i^w} \!\wedge\! \presence{a_i^h} \right) \!\!\neq\!\!
\left( \presence{a_{n+i}^w} \!\wedge\! \presence{a_{n+i}^h} \!\right) & i\in N \label{cp:3}
\\[-0.05cm]
&\presence{a_i^w} \neq \presence{a_{n+i}^w} & i\in N \label{cp:4}
\\[-0.05cm]
&\presence{a_i^h} \neq \presence{a_{n+i}^h} & i\in N \label{cp:5}
\\[-0.05cm]
\nonumber & \presence{a_i^w} \!\wedge\! \presence{a_i^h} \!\wedge\! \presence{a_j^w} \!\wedge\! \presence{a_j^h} \Rightarrow&
\\[-0.05cm]
\nonumber & \!\left(s_i^w + w_i \leq s_j^w\right) \!\vee\! \left(s_j^w + w_j \leq s_i^w\right) \!\vee\! \left(s_i^h + h_i \leq s_j^h\right) \!\vee\! \left(s_j^h + h_j \leq s_i^h\right)&
\\[-0.05cm]
&\qquad\qquad\qquad\qquad\qquad\qquad\qquad\qquad\qquad\qquad\qquad \left(i,j\right)\in {N^*}^2,\, i < j, \, j \neq n+ i& \label{cp:6}
\\[-0.05cm]
&\cumulative{\left[s_1^w,\ldots,s_{2n}^w\right]}{\left[w_1,\ldots,w_{2n}\right]}{H} & \label{cp:1}
\\[-0.05cm]
&\cumulative{\left[s_1^h,\ldots,s_{2n}^h\right]}{\left[h_1,\ldots,h_{2n}\right]}{W} & \label{cp:2}
\end{flalign}
}
Meta-constraint (\ref{cp:3}) guarantees that one of the pairs $(a_i^w,a_i^h)$ and $(a_{n+i}^w,a_{n+i}^h)$ is scheduled, where $(a_i^w,a_i^h)$ and $(a_{n+i}^w,a_{n+i}^h)$ correspond to item $i$ and its rotated duplicate $n+i$. It uses the ${\texttt{PresenceOf}}(a)$ constraint that signals the presence of optional activity $a,\ a \in A^w \cup A^h$. It returns \textit{true} when the optional activity $a$ is present, and \textit{false} otherwise.
Constraint (\ref{cp:4}) forbids scheduling activity $a_{n+i}^w$ when $a_i^w$ is scheduled and vice versa. Similarly, constraint (\ref{cp:5}) prohibits scheduling activity $a_{n+i}^h$ when $a_i^h$ is scheduled. Despite the presence of constraint (\ref{cp:3}), constraints (\ref{cp:4}) and (\ref{cp:5}) are needed to eliminate some infeasible cases.  For instance, constraint (\ref{cp:3}) discards neither the case where activities $a_i^w,\ a_i^h,$ and $a_{n+i}^w$ are scheduled while $a_{n+i}^h$ is not nor the case where activities $a_i^w,\ a_i^h,$ and $a_{n+i}^h$ are scheduled while $a_{n+i}^w$ is not.  On the other hand, constraints (\ref{cp:4}) and (\ref{cp:5}) remove these cases.  Constraint (\ref{cp:6}) ensures the no-overlap of any pair $(i,j) \in {N^*}^2$, $i < j$, $j \neq n+i$, of packed items. Its left hand side holds when activities $a_i^w$, $a_i^h$, $a_j^w$, and $a_j^h$ are scheduled and implies the right hand side, which is a disjunctive constraint that avoids the horizontal and vertical overlap of $i$ and $j$ by setting $i$ to the left of $j$ or $i$ above $j$ or $j$ to the left of $i$ or $j$ above $i$, where the \lq or\rq~ is inclusive. Finally, cumulative constraint (\ref{cp:1}) (resp. (\ref{cp:2})) makes the activities of $A^w$ (resp. $A^h$) complete within their respective time windows without exceeding the resource's capacity $H$ (resp. $W$). Even though redundant, constraints (\ref{cp:1})-(\ref{cp:2}) do strengthen the search.
The CP model returns a feasible solution if and only if every item of $N$ is assigned to the bin regardless of its rotation status.

For this model, the search tree is constructed as recommended by \citet{Clautiaux08}; i.e., variables $s_1^h,\ldots,s_{2n}^h$ are instantiated after variables $s_1^w,\ldots,s_{2n}^w$. The CP-model is solved via the \textsc{IBM ILOG CP Optimizer 12.6.2} ~\citep{Laborie2009}; set to the \textit{restart mode}, which applies a failure-directed search when its large neighborhood search fails to identify an improving solution ~\citep{Vilim}. That is, instead of searching for a solution, it focuses on eliminating assignments that are most likely to fail. (cf. \cite{DBLP:conf/flairs/LaborieR08} and \cite{Vilim2009} for basics of optional interval variables (i.e. optional activities) and cumulative constraints.)

When allocated a threshold run time {\TPACK}, the \textsl{CP Optimizer} acts as a heuristic, denoted hereafter as {\CPr}. Preliminary experiments showed that {\CPr} fathoms a large portion of infeasible solutions, especially when they are beyond \lq\lq the edge of feasibility\rq\rq.

%%%%%%%%%%%%%%%%%%%%%%%%%%%%%%%%%%%%%%%%%%%%%%%%%%%%%%%%%%%%%%%%%%%%%%%%%%%%%%%%%%%%%%%%%%%%%%%%%%%%%%%%%%%%%%%%%%
\section{Lower Bounds}\label{sec:LB}
%%%%%%%%%%%%%%%%%%%%%%%%%%%%%%%%%%%%%%%%%%%%%%%%%%%%%%%%%%%%%%%%%%%%%%%%%%%%%%%%%%%%%%%%%%%%%%%%%%%%%%%%%%%%%%%%%%
This section presents three lower bounds for {\BPPDD}: two existing and a new one. These three bounds are compared in Section \ref{ssec:CELB}. Herein, {\DMV} designates the linear-time lower bound algorithm of \citet{DellAmico200213} for the non-oriented two-dimensional bin packing problem while \DMVextra{S} is a lower bound on the number of bins needed to pack the items of set $S$.

%>>>>>>>>>>>>>>>>>>>>>>>>>>>>>>>>>>>>>>>>>>>>>>>>>>>>>>>>>>>>>>>>>>>>>>>>>>>>>>>>>>>>>>>>>>>>>>>>>>>>>>>>>>>>>>>>>
\subsection{Existing Lower Bounds}\label{LB1}
The procedure to calculate {\BSP} proceeds as follows. First, it sorts $N$ in a non-decreasing order of the due dates, and sets $[j]$ to the item with the $j^{\mbox{th}}$ earliest due date. It then uses \DMVextra{S_{[j]}}, $j=1,\ldots,n,$ to deduce a lower bound of the lateness of the subset of items $S_{[j]}=\{[1],\ldots,[j]\}$.  Some items must have a completion time $P \cdot$ \DMVextra{S_{[j]}}; thus, have a lateness of at least $P \cdot \text{\DMVextra{S_{[j]}}} - d_j.$ Therefore, {\BSP}$=\displaystyle\max_{j=1,\ldots,n} \{P \cdot \text{\DMVextra{S_{[j]}}} - d_j\}$ is a valid lower bound on $L_{max}$.

\citet{CLAUTIAUX2007365} use DFFs to compute lower bounds for the non-oriented bin packing problem when the bin is a square. They show that their bounds dominate {\DMV} both theoretically and computationally for square bins. However, for rectangular bins, the quality of this bound remains an open issue.

{\LBRELAX}, the second lower bound on $L_{max},$ is the result of the linear relaxation of {\MIP} where all the binary variables are substituted by variables in $\left[0,1\right].$

%>>>>>>>>>>>>>>>>>>>>>>>>>>>>>>>>>>>>>>>>>>>>>>>>>>>>>>>>>>>>>>>>>>>>>>>>>>>>>>>>>>>>>>>>>>>>>>>>>>>>>>>>>>>>>>>>>
\subsection{A New Lower Bound}\label{LB3}
To the opposite of {\LBRELAX}, which drops the integrality constraints, the new lower bound {\PH} is the \textbf{optimal} value of {\RMIP}, which is a mixed integer programming relaxation of {\MIP}. {\RMIP} exchanges the disjunctive constraints, given by Equations (\ref{eq:2})-(\ref{eq:6}), with the feasibility constraints
\begin{eqnarray}
&\displaystyle\sum_{i \in N} \left(\alpha^o_{ci} f_{ik} + \alpha^r_{ci} f_{n+ik} \right)\leq 1,\ k \in B,\ c=1,\ldots,m, \label{LB3:FC}
\end{eqnarray}
which are defined in the form of Equation (\ref{dff:FC}). The disjunctive constraints define the geometrical relationships between pairs of packed items and between a packed item and its assigned bin. They consider both the height and width of the items and bins and ensure the non-overlap of pairs of items and the containment of an item in the bin in both directions. The feasibility constraint, on the other hand, assimilates the item and the bin into dimensionless entities. Its inclusion in {\RMIP} tightens the relaxation and improves the quality of the lower bound. Excluding it omits the layout aspect of the problem; thus, can not produce reasonably good bounds.   {\PH} is a valid bound if and only if {\RMIP} is solved to optimality.

%%%%%%%%%%%%%%%%%%%%%%%%%%%%%%%%%%%%%%%%%%%%%%%%%%%%%%%%%%%%%%%%%%%%%%%%%%%%%%%%%%%%%%%%%%%%%%%%%%%%%%%%%%%%%%%%%%
\section{Approximate Approaches}\label{sec:UB}
%%%%%%%%%%%%%%%%%%%%%%%%%%%%%%%%%%%%%%%%%%%%%%%%%%%%%%%%%%%%%%%%%%%%%%%%%%%%%%%%%%%%%%%%%%%%%%%%%%%%%%%%%%%%%%%%%%
{\Alga} is a two-stage approximate approach for {\BPPDD}. The first stage constructs an initial solution and obtains related upper bounds using a new first-fit heuristic ({\FG}). The second stage is iterative. It improves the current solution via an assignment-based heuristic {\APr} and its relaxed version {\APrr}, and updates the bounds if possible. The second stage diversifies its search when it stagnates. Sections \ref{sec:FFC} - \ref{sec:SC} detail {\FG}, {\APr}, and {\Alga}.

%>>>>>>>>>>>>>>>>>>>>>>>>>>>>>>>>>>>>>>>>>>>>>>>>>>>>>>>>>>>>>>>>>>>>>>>>>>>>>>>>>>>>>>>>>>>>>>>>>>>>>>>>>>>>>>>>>
\subsection{First-Fit Heuristic}\label{sec:FFC}
{\FG} solves, via CP, a series of 2OPPs, where each 2OPP determines the feasibility of packing a given set of items into a single bin. It constructs a solution as detailed in Algorithm \ref{alg:1}. It sorts the items of $N$ in a non-descending order of their due dates, sets $k=0,$ and applies a sequential packing that iterates as follows until $N=\emptyset.$
First, it determines the current bin $k$ to be filled, and initializes its set $N_k$ of packed items to the empty set. It removes the first item from $N$ and inserts it into $N_k.$ \textbf{Two} scenarios are possible. \textbf{When} $\text{{\DMV}}\left(N_k\right) \leq 1,$ it considers the next item of $N.$ (The use of $\text{{\DMV}}\left(N_k\right)$ ensures that {\FG} starts with a dense packing; thus, limits the number of sequential calls to {\CPr}.) \textbf{Otherwise}, it undertakes a backward step followed by an iterative sequential packing step. It calls {\CPr} from Section \ref{sec:CPBA} to determine whether it is possible to pack the items of $N_k$. When infeasibility is detected (potentially because {\CPr} runs out of time), the backward step removes the last added item from $N_k$ (because it may have caused the infeasibility) and inserts $i$ back into $N.$ Then it calls {\CPr} again. When a feasible solution is obtained, {\FG} proceeds with the iterative sequential packing step.

The constructive step considers the items of $N$ sequentially. For every $i \in N,$ it checks whether a feasible packing is possible. Specifically, it calls {\CPr} when $\text{{\DMV}}\left(N_k \cup \left\{i\right\}\right) \leq 1.$ When {\CPr} determines that it is possible to pack the items of $N_k \cup \left\{i\right\}$ into the current bin, the constructive step removes $i$ from $N$ and inserts it into $N_k.$ Having tested all unpacked items of $N,$ {\FG} proceeds to the next bin by incrementing $k$ to $\left(k+1\right)$ if $N \neq \emptyset$.  Hence, {\FG} obtains an initial solution, characterized by its number of bins $b$ and its corresponding maximum lateness $\text{{\U}}.$  Subsequently, {\Alga} feeds this information to its second stage.

\setlength{\textfloatsep}{0.2cm}
\begin{algorithm}[t]
{\small
\caption{First-fit heuristic algorithm \textsc{FF}$\left(N\right)$}
\label{alg:1}
\begin{algorithmic}[1]
\State sort $N$ in a non-descending order of items' due dates;
\State set $k=0$;
\While{($N \neq \emptyset$)}
	\State set $k=k+1$;
	\State open a new bin $k$ setting $N_k = \emptyset$;
	\While{($\text{{\DMV}}\left(N_k \right) \leq 1$)}
		\State move the first item of $N$ to $N_k$;
	\EndWhile
	\While{({\CPr}$(N_k) \Longleftrightarrow infeasible)$}
		\State move the last item of $N_k$ back to its position in $N$;
	\EndWhile
	\For{each item $i \in N$} \label{code:1}
		\If{($\left( \text{\DMV} \left(N_k \cup \left\{i\right\}\right)\leq 1\right)$ \textbf{and} {\CPr}$(N_k \cup \left\{i\right\})\Longleftrightarrow feasible)$)} \label{code:3}
			\State move item $i$ from $N$ to $N_k$;
		\EndIf
	\EndFor \label{code:2}
\EndWhile
\State
\Return solution as $\left(N_1,\ldots,N_k\right)$;
\end{algorithmic}
}
\end{algorithm}

%>>>>>>>>>>>>>>>>>>>>>>>>>>>>>>>>>>>>>>>>>>>>>>>>>>>>>>>>>>>>>>>>>>>>>>>>>>>>>>>>>>>>>>>>>>>>>>>>>>>>>>>>>>>>>>>>>
\subsection{An Assignment-Based Heuristic} \label{sec:AP}
The second stage of {\Alga} applies iteratively an assignment-based heuristic {\APr}, which determines the feasibility of packing a set of oriented two-dimensional items into a set of multiple identical two-dimensional bins. Finding a feasible packing is hard not only because of the large number of alternative positions of an item within a bin but also because of the multitude of solutions having equal $L_{max}$. Herein, {\APr} implements \textbf{four} strategies that enhance its performance. \textbf{First}, it reduces the search space to a subset of feasible positions, which correspond to the free regions within a bin. As it applies its greedy search to position items, it creates some free regions and fills others; thus, the search space is dynamic. \textbf{Second}, {\APr} packs simultaneously as many items as possible into the various free regions. Thus, it reduces the number of iterations needed to obtain a solution; consequently, it decreases its runtime. \textbf{Third}, {\APr} implements a new kind of lookahead strategy that directs the search towards a feasible packing. This guiding mechanism imposes feasibility constraints that prohibit the current two-dimensional assignment problem {\ASSIGN} from generating partial solutions that will lead to infeasible ones in future iterations. This innovative mechanism makes current decisions account for their impact on future ones. \textbf{Fourth} and last, {\APr} uses upper bounds {\U} on $L_{max}$ and {\UbFunction{\U}} on the number of bins. These bounds further reduce the search space: a candidate solution is a feasible packing whose $L_{max} <$ {\U} and which uses at most {\UbFunction{\U}} bins. Initially, {\U} is the $L_{max}$ of the solution of {\FG}.

{\APr}, sketched in Algorithm \ref{alg:HEUR}, uses the following sets as input: the set $N$ of not yet packed items, the set $\overline{N}$ of packed items, and the set $E$ of available rectangular regions.  These three sets are updated dynamically at each iteration.  Initially, $N$ is the set of the $n$ items, $\overline{N} = \emptyset,$ and $E=B$, where $\left|B\right| = b = {\UbFunction{\U}}$. Thus, the set $E_k$ of free regions contained in bin $k,\ k \in B$, is initially the $k$th bin: $E_k=\{(W,H)\},$ with $E_k \subseteq E$ and $\displaystyle \cup_{k \in B} E_k = E$.

\begin{figure}[htb]
\centering
\includegraphics{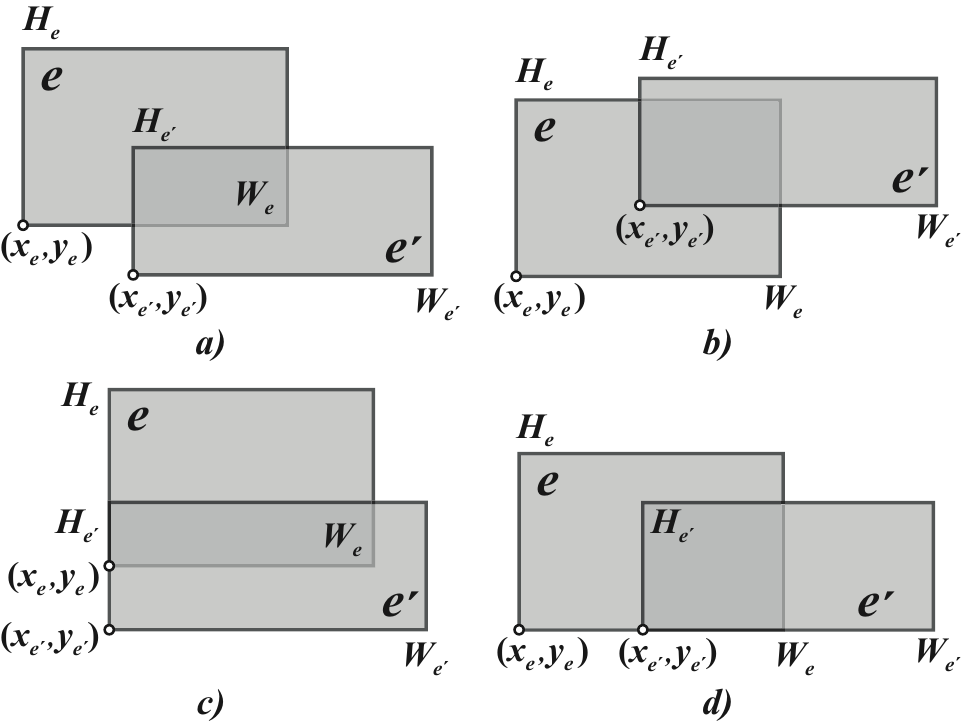}
\caption{Patterns where two regions $e$ and $e'$ overlap}\label{f1}
\end{figure}

Let $(e,e')\in E^2$ denote two free regions characterised by their respective dimensions $(W_e,H_e)$ and $(W_{e'},H_{e'})$ and by their bottom leftmost coordinate positions $(x_e,y_e)$ and $(x_{e'},y_{e'})$ in their respective bins. When in a same bin, $e$ and $e'$ may overlap, as in Figure \ref{f1}. To guard against assigning two items to the overlap area of $e$ and $e'$, {\APr} includes, into the assignment model, geometrical and disjunctive conditions that only apply if $e$ and $e'$ are in the same bin and overlap.  {\APr} signals such overlap via four parameters.
\begin{itemize}
\setlength\itemsep{-0.3em}
	\item $\thetav{1}=1$ if $\left(x_{e} < x_{e'}\right) \wedge \left(y_{e} > y_{e'}\right)$ as in Figure \ref{f1}.a and 0 otherwise.
	\item $\thetav{2}=1$ if $\left(x_{e} < x_{e'}\right) \wedge \left(y_{e} < y_{e'}\right)$ as in Figure \ref{f1}.b and 0 otherwise.
	\item $\thetav{3}=1$ if $\left(x_{e} = x_{e'}\right) \wedge \left(y_{e} > y_{e'}\right)$ as in Figure \ref{f1}.c and 0 otherwise.
	\item $\thetav{4}=1$ if $\left(x_{e} < x_{e'}\right) \wedge \left(y_{e} = y_{e'}\right)$ as in Figure \ref{f1}.d and 0 otherwise.
\end{itemize}
Similarly, it signals the already packed items via parameters
\begin{itemize}
\setlength\itemsep{-0.3em}
\item $\rho^o_{ik}=1$ if $i \in \overline{N}$ is packed in $k \in B$ without rotation and 0 otherwise; and
\item $\rho^r_{ik}=1$ if $i \in \overline{N}$ is packed in $k \in B$ with rotation and 0 otherwise.
\end{itemize}
When item $i$ is not yet packed (i.e., $i\in {N}$), $\displaystyle\sum_{k\in B} \rho^o_{ik}+\rho^r_{ik}=0$.

Let $E^U_i = \displaystyle\cup_{k \in B} \left\{E_k : kP - d_i < \text{{\U}} \right\}$ denote the set of regions where $i\in N$ can be scheduled and $L_i <$ {\U}. $E^U_i \supseteq E^o_i \cup E^r_i,$ where $E^o_i = \cup_{e \in E^U_i} \left\{e : \left(w_i \leq W_e\right) \wedge \left(h_i \leq H_e\right) \right\}$ and $E^r_i = \cup_{e \in E^U_i} \left\{e : \left(h_i \leq W_e\right) \wedge \left(w_i \leq H_e\right) \right\}$ are the sets of regions where $i$ can be positioned without and with rotation respectively.

In each iteration, {\APr} solves {\ASSIGN}, which attaches a subset of unpacked items to regions of $E$ using the following variables.
\begin{itemize}
\setlength\itemsep{0em}
\item $\varphi^o_{ie}=1$ if item $i\in N,$ is assigned without rotation to $e \in E^o_i$ and 0 otherwise.
\item $\varphi^r_{ie}=1$ if $i$ is assigned with rotation to $e \in E^r_i$ and 0 otherwise.
\item $f^o_{ik}=1$ if $i$ can be packed without rotation in a future iteration in bin $k$ such that $L_i=Pk - d_i < \text{{\U}},$ and 0 otherwise.
\item $f^r_{ik}=1$ if $i$ can be packed with rotation in a future iteration in bin $k$ such that $L_i=Pk - d_i < \text{{\U}},$ and 0 otherwise. $f^o_{ik}$ and $f^r_{ik}$ allocate free space for items to be packed in future iterations without increasing \text{{\U}}.
\item $w_e \in \left[0,W_e\right]$ and $h_e \in \left[0,H_e\right]$, the width and height of the used area of $e \in E$ when an item is positioned in $\left(x_e,y_e\right)$.
\item $l_{ee'}$ and $u_{ee'}$ are binary. They refer to the relative position of two areas $e$ and $e'.$ $l_{ee'}=1$ (resp. $u_{ee'}=1$) is used to make $e$ to the left of (resp. below) $e'$ when $e$ and $e'$ are part of the same bin.
\end{itemize}
{\ASSIGN} maximizes the total profit generated by the packed items subject to non-overlap and containment constraints. The profit of $i,\ i \in N,$ is $s_i \geq 0.$ When $s_i=w_i h_i,$ {\ASSIGN} maximizes the utilization of the bins; i.e., the density of the packed items. Formally, {\ASSIGN} is modeled as an {\MIPG}:

{\footnotesize
\begin{flalign}
\mbox{max}
& \sum_{i \in N} s_i \left(\sum_{e \in E^o_i} \frac{\varphi^o_{ie}}{W_eH_e} + \sum_{e \in E^r_i} \frac{\varphi^r_{ie}}{W_eH_e} \right)
&\label{eq:h0}
\\[-0.05cm]%------------------------------------------1
\mbox{s.t.}
& \sum_{\substack{i \in N: \\ e \in E^o_i}} \varphi^o_{ie} + \sum_{\substack{i \in N: \\ e \in E^r_i}} \varphi^r_{ie} \leq 1
&  e \in E \label{eq:h1}
\\[-0.05cm]%------------------------------------------2
& \sum_{e \in E^o_i} \varphi^o_{ie} + \sum_{e \in E^r_i} \varphi^r_{ie} + \spaceX\sum_{\substack{k \in B: \\ E_k \cap E^o_i \neq \emptyset}} \spaceX f^o_{ik} + \spaceX \sum_{\substack{k \in B: \\ E_k \cap E^r_i \neq \emptyset}} \spaceX f^r_{ik} = 1
&  i \in N \label{eq:h2}
\\[-0.05cm]%------------------------------------------3
\nonumber
& \sum_{e \in E_k} \left( \sum_{\substack{i \in N: \\ e \in E^o_i }} \alpha^o_{ci} \varphi^o_{ie} + \sum_{\substack{i \in N: \\ e \in E^r_i }} \alpha^r_{ci} \varphi^r_{ie} \right) + \spaceX \sum_{\substack{i \in N: \\ E_k \cap E^o_i \neq \emptyset}} \spaceX \alpha^o_{ci} f^o_{ik} +
&
\\[-0.05cm]%------------------------------------------4
& \sum_{\substack{i \in N: \\ E_k \cap E^r_i \neq \emptyset}} \spaceX \alpha^r_{ci}f^r_{ik} \leq 1 - \sum_{i \in \overline{N}} \left(\alpha^o_{ci}\rho^o_{ik} + \alpha^r_{ci}\rho^r_{ik}\right)
& k \in B, \, c=1,\ldots,m \label{eq:h3}
\\[-0.05cm]%------------------------------------------5
& \sum_{\substack{i \in N: \\ e \in E^o_i}} w_i\varphi^o_{ie} + \sum_{\substack{i \in N: \\ e \in E^r_i}} h_i\varphi^r_{ie} \leq w_e
&  e \in E \label{eq:h4}
\\[-0.05cm]%------------------------------------------6
& \sum_{\substack{i \in N: \\ e \in E^o_i}} h_i\varphi^o_{ie} + \sum_{\substack{i \in N: \\ e \in E^r_i}} w_i\varphi^r_{ie} \leq h_e
&  e \in E \label{eq:h5}
\\[-0.05cm]%------------------------------------------7
& x_e + w_e \leq x_{e'} + W_e \left(1 - l_{ee'}\right)
& (e,e') \!\in\! E^2 : \thetav{1} \!+\! \thetav{2} \!=\! 1 \label{eq:h6}
\\[-0.05cm]%------------------------------------------8
& y_{e'}+h_{e'}\leq y_e+H_{e'} \left(1-u_{e'e}\right)
& (e,e') \in E^2 : \thetav{1}=1 \label{eq:h7}
\\[-0.05cm]%------------------------------------------9
& y_e+h_e\leq y_{e'}+H_e\left(1-u_{ee'}\right)
& (e,e') \in E^2 : \thetav{2}=1 \label{eq:h13}
\\[-0.05cm]%------------------------------------------10
& h_{e'} + \left(y_{e'}+H_{e'}-y_e\right)\left(\sum_{\substack{i \in N: \\ e \in E^o_i}} \spaceS\varphi^o_{ie} + \spaceS\sum_{\substack{i \in N: \\ e \in E^r_i}} \spaceS\varphi^r_{ie}\right) \leq H_{e'}
& (e,e') \in E^2 : \thetav{3}=1 \label{eq:h15}
\\[-0.05cm]%------------------------------------------11
& w_e +\left(x_e+W_e-x_{e'}\right)\left(\sum_{\substack{i \in N: \\ e' \in E^o_i}} \spaceS\varphi^o_{ie'} +\spaceS\sum_{\substack{i \in N: \\ e' \in E^r_i}} \spaceS\varphi^r_{ie'}\right) \leq W_e
&  (e,e') \in E^2 : \thetav{4}=1 \label{eq:h16}
\\[-0.05cm]%------------------------------------------12
& l_{ee'}+u_{e'e} \geq 1
& (e,e') \in E^2 : \thetav{1}=1 \label{eq:h8}
\\[-0.05cm]%------------------------------------------13
& l_{ee'}+u_{ee'} \geq 1
& (e,e') \in E^2 : \thetav{2}=1 \label{eq:h14}
\\[-0.05cm]%------------------------------------------14
& \varphi^o_{ie} \in \left\{0,1\right\}, \; \varphi^r_{ie} \in \left\{0,1\right\}
& i \in N, \;e \in E \label{eq:h9}
\\[-0.05cm]%------------------------------------------15
& f^o_{ik} \in \left\{0,1\right\}
& i \!\in\! N, \; k \!\in\! B:E_k \!\cap\! E^o_i \!\neq\! \emptyset \label{eq:h10}
\\[-0.05cm]%------------------------------------------16
& f^r_{ik} \in \left\{0,1\right\}
& i \!\in\! N, \; k \!\in\! B:E_k \!\cap\! E^r_i \!\neq\! \emptyset \label{eq:h10b}
\\[-0.05cm]%------------------------------------------17
\nonumber
& l_{ee'}\in \left\{0,1\right\}, \; u_{ee'} \in \left\{0,1\right\}
& (e,e') \in E^2 :
\\[-0.05cm]%------------------------------------------18
&
& \left(\thetav{1} = 1\right) \vee \left(\thetav{2} = 1\right) \label{eq:h11}
\\[-0.05cm]%------------------------------------------
& 0 \leq w_e \leq W_e, \; 0 \leq h_e \leq H_e
& e \in E
\label{eq:h12}
\end{flalign}}

Equation (\ref{eq:h0}) defines the objective function value as the weighted sum of the profits of packed items where the weight of an item $i$ is inversely proportional to the area of region $e$ used for its positioning. Equation (\ref{eq:h1}) prohibits packing more than one item into any region $e \in E$.

Equations (\ref{eq:h2}) and (\ref{eq:h3}) are part of the lookahead strategy. They employ $f^{*}_{ik},\ *=o,\,r,\ i \in N,\ k \in B$, to reserve a free space for unpacked items. Equation (\ref{eq:h2}) assigns $i,\ i \in N$, either to one of the available regions during the current iteration or to one of the bins during a later iteration. Equation (\ref{eq:h3}) imposes the set of feasibility constraints. Here, $c,\ c=1,\ldots,m$ determines a vector of transformed areas ($\alpha^o_{ci}$ and $\alpha^r_{ci}$) computed for all the items on $N \cup \overline{N}$ and their rotated copies and represented via matrices $\mathcal{A}^o$ and $\mathcal{A}^r$ (cf. Section \ref{sec:fc}). For every $c,\ c=1,\ldots,m,$ and $k \in B$, Equation (\ref{eq:h3}) requires that the sum of the transformed areas of (i) the items that have been previously packed ($\rho^{*}_{ik}=1,\ *=o,\ r$), (ii) those being packed at the current
iteration ($\varphi^{*}_{ie}=1,\ *=o,\ r$), and (iii) those to be packed in future iterations ($f^{*}_{ik}=1,\ *=o,\ r$) in selected bin $k$ be bounded by 1. Even though it discards many partial solutions that lead to an infeasible packing, Equation (\ref{eq:h3}) doesn't guarantee that a not-yet-packed $i$ will get a feasible position during later iterations.

Equations (\ref{eq:h4}) and (\ref{eq:h5}) determine $w_e$ and $h_e$ of the used area of $e$ by imposing that $w_i$ and $h_i$ do not exceed $w_e$ and $h_e$ if $i$ is assigned to $e$.

Equations (\ref{eq:h6})-(\ref{eq:h14}) guarantee the non-overlap of a pair of items packed in two overlapping regions $(e,e')$. They substitute the full set of the disjunctive constraints that are traditionally used to ensure the non-overlap of packed items in a bin. This substitution reduces the number of constraints by eliminating redundant ones. That is, instead of considering all possible pairs of regions, {\ASSIGN} focuses on those that can potentially create an overlap of packed items. It detects these regions via parameters $\thetav{1}$ to $\thetav{4}$.

Equations (\ref{eq:h6})-(\ref{eq:h13}) focus on the case where $e$ is to the left of $e'$ but $e$ and $e'$ overlap. For those regions, Equation (\ref{eq:h6}) makes the $x-$coordinate of the rightmost point of the used area of $e$ less than or equal to its counterpart for the leftmost point of $e'$. Equations (\ref{eq:h7}) and (\ref{eq:h13}) constrain the vertical positions of the used areas of $e$ and $e'$. Equation (\ref{eq:h7}) deals with the case when $e'$ is below $e$ and $\thetav{1} = 1$ as in Figure \ref{f1}.a. It restricts the $y-$coordinate of the topmost point of the used area of $e'$ to be less than or equal to its counterpart of the bottommost point of $e$. Similarly, when $e'$ is below $e$ and $\thetav{2}= 1$, Equation (\ref{eq:h13}) constrains the topmost $y-$coordinate of the used area of region $e$ to be less than or equal to the lowest $y-$coordinate of region $e'$; thus avoiding the potential overlap of items assigned to the two regions depicted in Figure \ref{f1}.b.

Equations (\ref{eq:h15}) and (\ref{eq:h16}) deal with two special cases: the left sides of $e$ and $e'$ are aligned vertically, and the bottom sides of $e$ and $e'$ are aligned horizontally. When $e$ and $e'$ are aligned vertically and an item is packed in $e$, Equation (\ref{eq:h15}) constrains the topmost $y-$coordinate of the used area of region $e'$ to be less than or equal to the lowest $y-$coordinate of region $e$ as in Figure \ref{f1}.c. Similarly, when both $e$ and $e'$ are aligned horizontally and an item is positioned into $e'$, Equation (\ref{eq:h16}) restricts the rightmost $x-$coordinate of the used area of region $e$ to be less than or equal to the leftmost $x-$coordinate of $e'$ as in Figure \ref{f1}.d.

Equations (\ref{eq:h8}) and (\ref{eq:h14}) ensure that the used areas of any pair of overlapping regions $(e,e')$ are such that the used area of $e'$ is below $e$, the used area of $e$ is below $e'$ or the used area of $e$ is to the left of $e'$.

Finally, Equations (\ref{eq:h9})-(\ref{eq:h12}) declare the types of the decision variables.

\begin{algorithm}[t]
{\small
\caption{Assignment-Based Heuristic {\APr}({\U})} \label{alg:HEUR}
\begin{algorithmic}[1]
\State initialize $\overline{N} = \emptyset$ and $E=B$, $\left|B\right| = b = {\UbFunction{\U}}$;
\While{(true)}
	\If{($\ASSIGN\left(N,\overline{N},E\right) \Longleftrightarrow feasible$)}
		\For{each item $i \in N$}
			\If{($\exists \ e\in E: \left(\varphi^o_{ie}=1\right) \vee \left(\varphi^r_{ie}=1\right)$)}
				\State set $\rho^{o}_{ik}=\varphi^o_{ie}$ and $\rho^{r}_{ik}=\varphi^r_{ie}$ for $k:e \in E_k$;
				\State move item $i$ from $N$ to $\overline{N}$;
			\EndIf
		\EndFor	
		\If{($N=\emptyset$)}
			\State return \textit{feasible solution};
		\EndIf
		\State calculate coordinates $\left(x_i',y_i'\right)$ and $\left(x_i'',y_i''\right)$ for every item $i \in \overline{N}$;
		\State update $E$ exploring regions on top and to the right of every item $i \in \overline{N}$;
		\For{each region $e \!\in\! E_k\!$, $k \!\in\! B$}
			\If{($\nexists \ i\!\in\! N\!:\! \left(w_i\!\leq\! W_e\right) \!\wedge\! \left(h_i\!\leq\! H_e\right) \!\wedge\! \left(kP\!-\!d_i\!<\!\U\right)$)}
				\State move $e$ from $E$ to $\overline{N}$;
			\EndIf
		\EndFor	
		\If{($\exists \ i\in N: E^o_i \cup E^r_i=\emptyset$)}
			\State return \textit{infeasible solution};
		\EndIf
	\Else
		\State return \textit{infeasible solution};
	\EndIf
\EndWhile
\end{algorithmic}
}
\end{algorithm}

When {\ASSIGN} returns a feasible solution, {\APr} moves the packed items from $N$ to $\overline{N}$, and sets the parameters $\rho^{*}_{ik}=1,\ *=o,\ r, \ i \in \overline{N},\ k \in B$, of the items packed in the current iteration. Next, it calculates the coordinates $\left(x_i',y_i'\right)$ and $\left(x_i'',y_i''\right)$ of both the upper left and the bottom right corners of item $i \in \overline{N}$, where $\left(x_i',y_i'\right) = \left(x_i,y_i + h_i\rho^o_{ik} + w_i\rho^r_{ik}\right)$ and $\left(x_i'',y_i''\right) = \left(x_i + w_i\rho^o_{ik} + h_i\rho^r_{ik},y_i\right)$. Finally, {\APr} updates $E$ using the following two-step approach.

\begin{figure}[tb]
\centering
\includegraphics{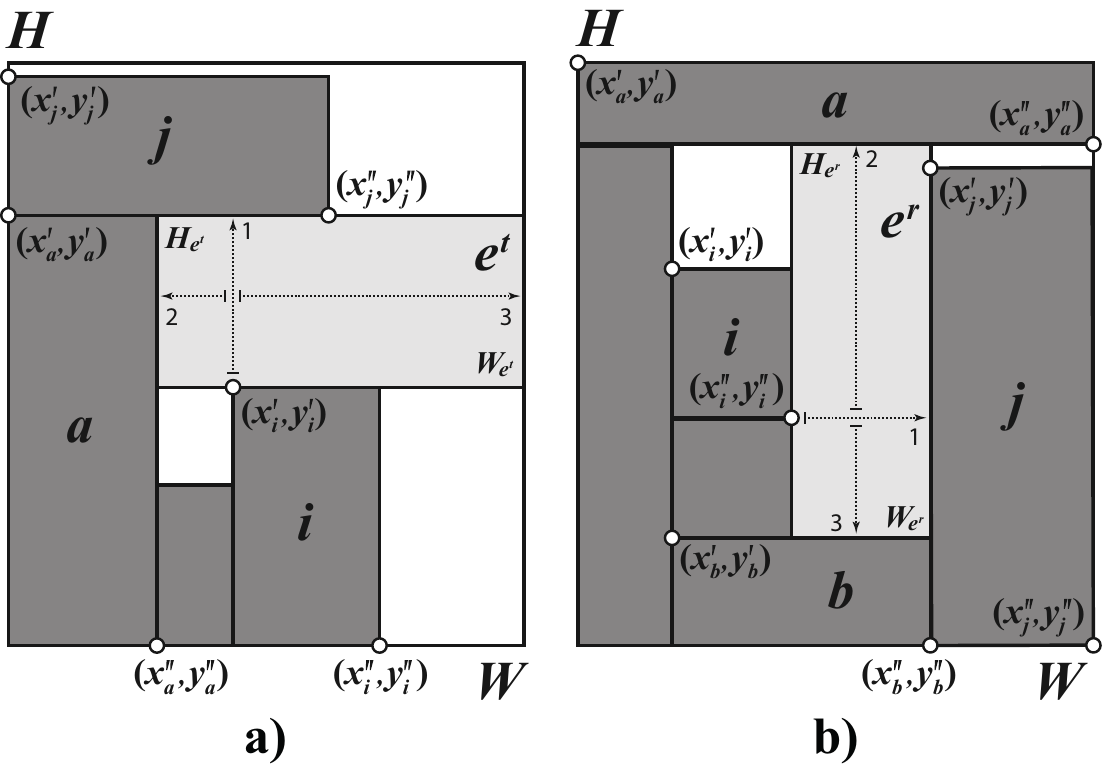}
\caption{New regions created by the partial packing of a bin}
\label{f2}
\end{figure}

The first step defines the region $e^t=(W_{e^t},H_{e^t})$ on top of item $i$ (cf. Figure \ref{f2}.a). To identify the height $H_{e^t}$, it searches along the ray $x=x_i$ and $y \geq y_i'$ for the first bottom side of another item $j$ if such an item exists or the upper side of the bin. It sets $H_{e^t} = y^t-y_i',$ where $y=y^t$ is the line intersecting this side, with $y^t=y_{j}$ if $j$ exists, and $y^t=H$ otherwise. To determine the width $W_{e^t}$, it expands its search along the line $y=y_i'$; i.e., to both the left and right sides of $x=x_i.$ It shifts the left edge of $e^t$ until it meets the first right edge of an item $a$ or the left border of the bin. It determines the line $x=x^{\ell}$ intersecting this side where $x^{\ell}=x_a''$ if $a$ exists, and $x^{\ell}=0$ otherwise. Similarly, it moves the right edge of $e^t$ until it meets either the first left edge of an item $b$ or the right border of the bin. It finds the line $x=x^r$ intersecting this side where $x^r=x_{b}$ if $b$ exists, and $x^r=W$ otherwise. Subsequently, $W_{e^t}=x^r-x^l$.

The second step defines the region $e^r=(W_{e^r},H_{e^r})$ to the right of $i$ (cf. Figure \ref{f2}.b). It identifies the width $W_{e^r}$ by searching along the ray $y=y_i,\ x \geq x_i''$ for the first left side of another item $j$ if $j$ exists or the right side of the bin. It finds the line $x=x^r$ intersecting this side where $x^r=x_{j}$ if $j$ exists and $x^r=W$ otherwise. It deduces $W_{e^r} = x^r-x_i''$. It determine the height $H_{e^r}$ by expanding its search along the line $x=x_i''$; i.e., above and below $y=y_i.$ It moves the top edge of $e^r$ until it meets either the first bottom edge of an item $a$ or the top border of the bin. It sets the line intersecting this side to $y=y^t$ where $y^t=y_a''$ if $a$ exists, and $y^t=H$ otherwise. Similarly, it shifts the bottom edge of $e^r$ until it meets the first top edge of an item $b$ or the lower border of the bin. It finds the line $y=y^b$ intersecting this side where $y^b=y_b'$ if $b$ exists, and $y^b=0$ otherwise.  Subsequently, $H_{e^r}=y^t-y^b$.

Having defined $e^t$ and $e^r$, {\APr} examines their \lq\lq utility\rq\rq. It discards $e$ if $e$ cannot hold at least one of the unpacked items of $N$ or $e$ can hold an unpacked item but yields a lateness that is larger than or equal to {\U}. {\APr} inserts a discarded region $e$ into $\overline{N}$ treating $e$ as a dummy item packed in bin $k$. This insertion strengthens Equation (\ref{eq:h3}). Finally, {\APr} inserts all non-discarded regions into $E$, and checks whether the stopping criterion is satisfied. In fact, {\APr} stops when all the items are packed or there is an unpacked item that does not fit into any region of $E$. When the stopping criterion is not met, {\APr} runs another iteration with the updated $N$ and $\overline{N}$.

%>>>>>>>>>>>>>>>>>>>>>>>>>>>>>>>>>>>>>>>>>>>>>>>>>>>>>>>>>>>>>>>>>>>>>>>>>>>>>>>>>>>>>>>>>>>>>>>>>>>>>>>>>>>>>>>>>
\subsection{Solution Process as a Whole}\label{sec:SC}

\begin{algorithm}[t]
{\small
\caption{The framework of {\Alga}} \label{alg:2}
\begin{algorithmic}[1]
\State \textbf{Stage 1: Initialization}
\State run {\FG} to obtain an initial solution; \label{step:1}
\State compute {\U} and $b \!=\! {\UbFunction{\U}} \!=\! \displaystyle \max_{i \in N} \left\{\!\left\lfloor\left(\!\text{{\U}} \!+\! d_i\right) \!/\! P\right\rfloor \! \right\}$ based on the solution of Line \ref{step:1};
\State \textbf{Stage 2: Iterative Step}
\Repeat  \label{step:3}
	\State set $s_i=w_i h_i, \ i \in N$;
	\State set ${\co}=0$;
	\Repeat \label{step:5}
		\If{(call $\APrr\left(\U\right) \Longleftrightarrow feasible$)} \label{step:2}
				\State compute {\U} and $b= {\UbFunction{\U}}$ based on the solution of Line \ref{step:2};
				\State break;		
		\Else
				\State modify $s_i=\gamma w_i h_i,\ i \in N,$ in Equation (\ref{eq:h0}) of {\ASSIGN};
				\State set ${\co}={\co}+1$;
		\EndIf
	\Until{${\co} \leq \text{\LHEUR}$} \label{step:6}
	\If{${\co} > \text{\LHEUR}$}
		\State break;
	\EndIf
\Until{true} \label{step:4}
\State repeat lines \ref{step:3}-\ref{step:4} replacing {\APrr} with {\APr};
\end{algorithmic}
}
\end{algorithm}

{\Alga}, detailed in Algorithm \ref{alg:2}, consists of two stages. The first stage applies {\FG} to obtain an initial feasible solution to {\BPPDD} along with an upper bound {\U} on $L_{max}$ and an upper bound {\UbFunction{\U}} on the number of bins in an optimal solution.  The second stage strives to improve these two bounds and the current solution using both the assignment-based heuristic {\APr} and its relaxed version {\APrr}. Specifically, it solves the non-oriented 2OPP with $b= {\UbFunction{\U}} = \displaystyle \max_{i \in N} \left\{\left\lfloor\left(\text{{\U}} + d_i\right) / P\right\rfloor\right\}$ bins so that the maximal lateness of any feasible solution is less than {\U}. It solves the problem in two steps, each consisting of two loops: An outer loop whose objective is to identify a solution with a tighter {\U} rapidly and an inner loop whose objective is to refine the search.

In the first step, the outer loop (cf. lines \ref{step:3}-\ref{step:4} of Algorithm \ref{alg:2}) resets the profits $s_i=w_i h_i, \ i \in N$, and the iteration counter {\co} to 1. Then the inner loop runs {\APrr} (cf. lines \ref{step:5}-\ref{step:6}), which is a reduced version of {\APr} where Equation (\ref{eq:h3}) and the decision variables $f^o_{ik}$ and $f^r_{ik},\ i \in N, \ k \in B$, are omitted from the model {\ASSIGN}. {\APrr} is generally weaker than {\APr} in terms of the tightness of the upper bound of lateness but is faster in terms of run time. When it obtains a feasible solution, {\APrr} feeds {\Alga} with a solution whose $L_{max} <$ {\U}; that is, it tightens {\U}. This feasible solution may also reduce $b={\UbFunction{\U}}$. Subsequently, {\Alga} exits the inner loop and runs one more iteration calling {\APrr} again but this time with new values of {\U} and $b$.

On the other hand, when {\APrr} fails to find a feasible solution, the inner loop diversifies the search by using a different set of random profits. It changes the profits to $s_i=\gamma w_i h_i, \ i \in N,$ where $\gamma$ is a random real from the continuous \textsl{Uniform}[1,3], and increments {\co} by 1. If {\co} is less than or equal to a maximal number of iterations {\LHEUR}, the inner loop starts a new iteration by solving {\APrr} with its modified profits in the objective function (i.e., in Equation (\ref{eq:h0}) of {\ASSIGN}).

When {\co} reaches the limit {\LHEUR}, {\Alga} proceeds with the second step, which performs exactly the same actions as the first step does except that it applies {\APr} instead of {\APrr}. The use of {\APr} should improve the search. Therefore, the first step pre-solves the problem quickly while the second looks for an enhanced solution.

Modifying the weight coefficients of Equation (\ref{eq:h0}) of {\ASSIGN} is a random local search ({\RLS}). The choice of this particular diversification strategy along with this specific range of $\gamma$ was based on preliminary computational investigations. Tests have shown that {\RLS} yields, on average, better results than evolutionary strategies and techniques such as the method of sequential value correction \citep{Belov08}. The superiority of {\RLS} is due to the items' random order, which is further accentuated by the unequal weights. Classical approaches on the other hand do not tackle the highly symmetric nature of bin packing solutions. They mainly construct solutions based on the sequential packing of items in ascending order of their areas/widths/heights \citep{Lodi2002379}.

%%%%%%%%%%%%%%%%%%%%%%%%%%%%%%%%%%%%%%%%%%%%%%%%%%%%%%%%%%%%%%%%%%%%%%%%%%%%%%%%%%%%%%%%%%%%%%%%%%%%%%%%%%%%%%%%%%
\section{Computational Experiments}\label{sec:CE}
%%%%%%%%%%%%%%%%%%%%%%%%%%%%%%%%%%%%%%%%%%%%%%%%%%%%%%%%%%%%%%%%%%%%%%%%%%%%%%%%%%%%%%%%%%%%%%%%%%%%%%%%%%%%%%%%%%

The objective of the computational investigation is fourfold. First, it compares the proposed lower bound {\PH} to both {\BSP} and {\LBRELAX}.  Second, it assesses the quality of the solution values of {\FG}, {\Alga} and {\MIP}. Third, it compares the performance of {\FG} and {\Alga} to that of {\GA}. Last, it evaluates the performance of {\FG} and {\Alga} on large-sized instances. All comparisons apply the appropriate statistical tests. All inferences are made at a 5\% significance level, and all confidence interval estimates have a 95\% confidence level.

{\Alga} is implemented in C\#, which evokes \textsl{IBM ILOG Optimization Studio 12.6.2} to handle {\MIPG} and CP models. It is run on a PC with a 4 Gb RAM and a 3.06 GHz Dual Core processor. The time limit {\TPACK} for {\CPr} is set to 2 seconds. This setting, inferred from preliminary computational investigations, gives the best tradeoff between density of packing and runtime. Indeed, a longer {\TPACK} does not necessarily lead to better packing solutions while it unduely increases the runtime of {\FG}.  Similarly, a shorter {\TPACK} often hinders {\FG} from reaching a feasible packing; thus causes poor quality solutions. The maximal number of iterations for {\co} is $\text{\LHEUR}=100$, which also represents the best trade-off between quality and performance of {\Alga} according to our earlier tests. Furthermore, $p\in\left\{0.15,\,0.3,\,0.45\right\}$, $q\in\left\{0.15,\,0.3,\,0.45\right\}$, and up to $m=27$ feasibility constraints are generated for the model {\ASSIGN} of Section \ref{sec:AP}. A larger number of constraints does not generally improve the solution quality but increases the runtime of {\ASSIGN}.  Despite their large variety, the feasibility constraints of {\ASSIGN} do not always tighten the lower bound on the free space available for packing. Therefore, their larger number does not necessarily tighten the model.

Section \ref{ssec:CEsetup} presents the benchmark set. Section \ref{ssec:CELB} measures the tightness of $LB_3.$ Section \ref{ssec:CEUB} assesses the performance of {\FG}, {\Alga} and {\MIP} in terms of their optimality gaps and number of times they reach the optimum. Section \ref{ssec:CEGA} compares the results of {\FG}, {\Alga} and {\GA}.  Finally, Section \ref{ssec:scalability} studies the sensitivity of {\FG}, {\Alga} and {\MIP} to problem size.

%>>>>>>>>>>>>>>>>>>>>>>>>>>>>>>>>>>>>>>>>>>>>>>>>>>>>>>>>>>>>>>>>>>>>>>>>>>>>>>>>>>>>>>>>>>>>>>>>>>>>>>>>>>>>>>>>>
\subsection{Computational Set Up}\label{ssec:CEsetup}

\citet{Bennell13} generated the benchmark set (including the due dates) that we test. For each instance, they calculated {\BSP}, and applied their multi-crossover genetic algorithm {\GA} to obtain upper bounds. ({\GA} was coded in ANSI-C using Microsoft Visual C++ 6.0 and run on a Pentium 4, 2.0 GHz, 2.0 GB RAM computer with a 120-second time limit per replication, and ten replications per instance.)  They, then, computed the average percent deviations of their upper bounds from {\BSP}. In their paper, they reported these average deviations aggregated over problem size. We use their aggregated average deviations in the comparisons of Section~\ref{ssec:CEGA}.  However, we recomputed their {\BSP} for every instance to perform the comparisons of Sections~\ref{ssec:CELB} and~\ref{ssec:CEGA}.

Their benchmark set uses square bins ($W=H$) whose processing times $P=100.$ It consists of 10 categories as detailed in Table \ref{tab:type}. Column 3 gives the width $W$ of a bin. Column 4 specifies how items are generated. Each category is characterised by the dimensions of the items, with categories 1-6 having homogeneous items that are randomly generated from a specific discrete uniform whereas categories 7-10 contain heterogeneous items belonging to four types in various proportions. The four types correspond to items whose $(w_i,h_i)$ are randomly selected from discrete uniforms on the respective ranges:
\begin{itemize}
\setlength\itemsep{0em}
	\item type 1: $\left(\left[\frac{2}{3}W,W\right],\left[1,\frac{1}{2}W\right]\right)$;
	\item type 2: $\left(\left[1,\frac{1}{2}W\right],\left[\frac{2}{3}W,W\right]\right)$;
	\item type 3: $\left(\left[\frac{1}{2}W,W\right],\left[\frac{1}{2}W,W\right]\right)$; and
	\item type 4: $\left(\left[1,\frac{1}{2}W\right],\left[1,\frac{1}{2}W\right]\right)$.
\end{itemize}
The categories can be divided, according to the relative size of the items, into two sets $\mathcal{L}$ and $\mathcal{S}$. Set $\mathcal{L},$ which contains instances with relatively large items, consists of categories 1, 3, 5, 7, 8, and 9. Set $\mathcal{S}$, which contains instances with small items, consists of categories 2, 4, 6, and 10.

\begin{table}[htb]
\centering
\caption{Generation of the widths and heights of items}\label{tab:type}
{\scriptsize
\begin{tabular}{rrrl}
Category&Set&Bin size ($W$)& Item size $\left(w_i,h_i\right)$\\
\hline
1&$\mathcal{L}$&10&uniformly random in $\left[1,10\right]$\\
2&$\mathcal{S}$&30&uniformly random in $\left[1,10\right]$\\
3&$\mathcal{L}$&40&uniformly random in $\left[1,35\right]$\\
4&$\mathcal{S}$&100&uniformly random in $\left[1,35\right]$\\
5&$\mathcal{L}$&100&uniformly random in $\left[1,100\right]$\\
6&$\mathcal{S}$&300&uniformly random in $\left[1,100\right]$\\
7&$\mathcal{L}$&100&type 1 with probability 70\%; type 2, 3, 4 with probability 10\% each\\
8&$\mathcal{L}$&100&type 2 with probability 70\%; type 1, 3, 4 with probability 10\% each\\
9&$\mathcal{L}$&100&type 3 with probability 70\%; type 1, 2, 4 with probability 10\% each\\
10&$\mathcal{S}$&100&type 4 with probability 70\%; type 1, 2, 3 with probability 10\% each\\
\end{tabular}
}
\end{table}

For each category, there are five problem sizes: $n= 20,\ 40,\ 60,\ 80,$ and $100,$ and ten instances per category and problem size.  For each problem, there are three classes A, B, and C of due dates, generated from the discrete \textsl{Uniform}$[101, \beta P\cdot\text{\DMV}]$ where $\beta = 0.6, 0.8,$ and $1.0;$ thus, a total of 1500 instances.

%>>>>>>>>>>>>>>>>>>>>>>>>>>>>>>>>>>>>>>>>>>>>>>>>>>>>>>>>>>>>>>>>>>>>>>>>>>>>>>>>>>>>>>>>>>>>>>>>>>>>>>>>>>>>>>>>>
\subsection{Quality of the Lower Bounds}\label{ssec:CELB}

This section compares the performance of {\BSP}, {\LBRELAX}, and {\PH}, where {\BSP} is computed via the algorithm of Section \ref{LB1}, {\LBRELAX} is the value of the incumbent returned by \textsl{CPLEX} and {\PH} is the optimal value of {\RMIP} when \textsl{CPLEX} identifies the optimum within 1 hour of runtime. Table \ref{Tab:LB} summarizes the statistics of the lower bounds per class, category, and problem size.  It displays
\begin{itemize}
\setlength\itemsep{0em}
\item $\gamma_{\bullet},$ the percent deviation of $LB_{\bullet},\ \bullet=1,2,3,$ from the tightest lower bound $LB^*$ where $LB^*=\max\{ \BSP,\LBRELAX,\PH, \MIP \},$ and $\gamma_{\bullet}=100 {(LB^*-LB_{\bullet})}/{LB^*},$ with {\MIP} included in the computation of $LB^*$ only when \MIP~ is proven optimal;
\item $\eta_{\bullet},$ the number of times $LB_{\bullet}=LB^*,\ \bullet = 1,2,3;$ and
\item $\#,$ the number of times {\PH} is not a valid bound; i.e., the number of times the linear programming solver \textsl{CPLEX} fails to prove the optimality of its incumbent within the 1 hour time limit.
\end{itemize}

Table \ref{Tab:LBRT} reports statistics of the runtime of {\PH} along with the tallied $\#$ per class, category and problem size. The statistics of the runtime are the average $RT$, median $Q_2$, minimum $\underline{RT}$ and maximum $\overbar{RT};$ all in seconds. The median (i.e., the 50th percentile) separates the ordered data into two parts with equal numbers of observations. It is a more appropriate measure of central tendency in the presence of outliers or when the distribution of the data is skewed.

The \textbf{Category} rows of Tables \ref{Tab:LB} and \ref{Tab:LBRT} display the same statistics as the tables but per category per class. Their last rows report these statistics per class.  Finally, their last eight columns give the statistics over all classes. A missing value indicates that all ten instances are unsolved by {\PH}; i.e., \# = 10.

The analysis of Tables \ref{Tab:LB} and \ref{Tab:LBRT} suggests the following.  {\BSP} is the best lower bound in 1083 instances out of 1500 instances. Over all instances, its average deviation from $LB^*$ is 5.7\%.  Its runtime is very reduced.

{\LBRELAX} never outperforms {\PH} nor {\BSP}.  It matches $LB^*$ for only 111 instances; i.e., in 7.40\% of the cases.  These instances have $n=20$ and $40$, and belong to categories 2, 4, and 6. The average percent deviation of {\LBRELAX} from  $LB^*$ is 72.5\%.

{\PH} is a valid bound for 1244 instances. Its average runtime is 74.91 seconds. Its much smaller median (2.65 seconds) signals the existence of some outlier cases that increased the mean. This is expected since \textsl{CPLEX} is allocated up to one hour to prove the optimality of its incumbent.

For those 1244 instances, {\PH} enhances {\BSP} for 361 out of 1500 instances; i.e., in 24.07\% cases. Its average enhancement over these 361 instances is 31.30\%.  In addition, it matches {\BSP} for another 463 instances; i.e., in 30.87\% cases.  Subsequently, it is the best lower bound (among {\PH}, {\LBRELAX}, {\BSP}) in 824 cases.

\begin{figure}[htb]
\centering
\includegraphics[width=\textwidth]{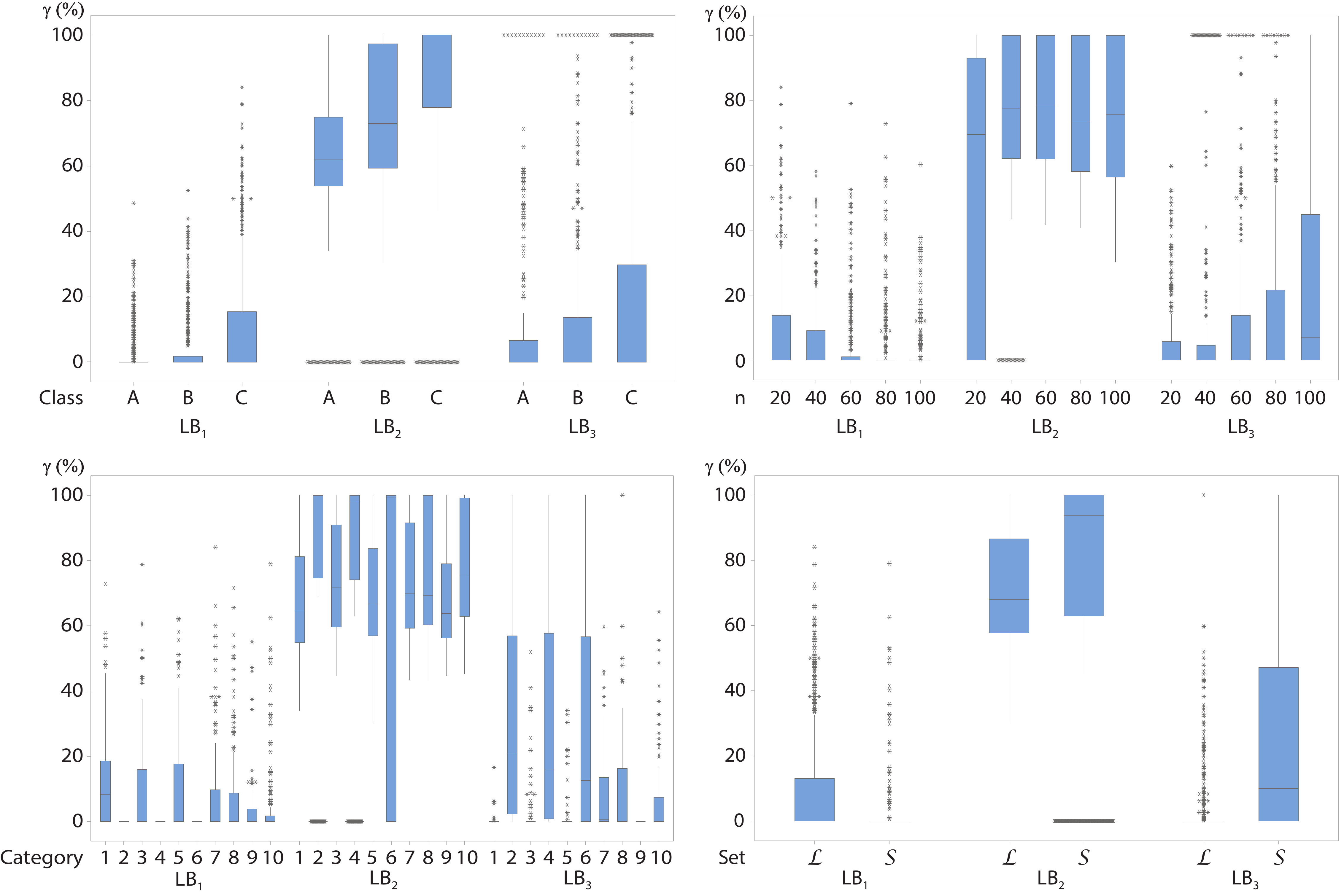}
\caption{Box plots for the mean percent deviations of the lower bounds from $LB^*$}\label{Fig:LB}
\end{figure}

Figure \ref{Fig:LB} displays the box plots of the percent deviations $\gamma_{\bullet}$ of $LB_{\bullet},\ \bullet=1,2,3,$ from $LB^*$ as a function of the class, size, category and set of the instances.  A box plot reflects the central tendency, spread, and skewness of the observed values.  Its box corresponds to the 25th, 50th and 75th percentiles whereas its fences extend to the lowest and largest value of the data.  Its stars signal outliers or unusual observations. Figure \ref{Fig:LB} infers that {\PH} is mostly superior for set $\mathcal{L}$; that is, for all sizes of categories 1, 3, 5, 9 and for small-sized instances ($n=20$ and $40$) of categories 7 and 8. Overall, the average percent deviations of {\PH} from {\BSP} and from $LB^*$ are 3.58\% and 14.1\%. Furthermore, {\PH} strictly dominates {\LBRELAX} in 1074 cases. The three quartiles of the percent improvement, over all instances, are: 146.4, 218.9 and 417.1; implying a larger enhancement for the cases with strict dominance.  That is, {\PH} is at least one order of magnitude larger than {\LBRELAX} in most instances.

\begin{sidewaystable}
\caption{Comparison of lower bounds}\label{Tab:LB}
\vskip+0.5cm
\begin{tiny}{
\hskip-1.0cm
\hspace*{1cm}
\begin{tabular}{@{\,\,\,\,}r@{\,\,\,\,}r@{\,\,\,\,}||
@{\,\,\,\,}r@{\,\,\,\,}r@{\,\,\,\,}r@{\,\,\,\,}|
@{\,\,\,\,}r@{\,\,\,\,}r@{\,\,\,\,}r@{\,\,\,\,}r@{\,\,\,\,}||
@{\,\,\,\,}r@{\,\,\,\,}r@{\,\,\,\,}r@{\,\,\,\,}|
@{\,\,\,\,}r@{\,\,\,\,}r@{\,\,\,\,}r@{\,\,\,\,}r@{\,\,\,\,}||
@{\,\,\,\,}r@{\,\,\,\,}r@{\,\,\,\,}r@{\,\,\,\,}|
@{\,\,\,\,}r@{\,\,\,\,}r@{\,\,\,\,}r@{\,\,\,\,}r@{\,\,\,\,}||
@{\,\,\,\,}r@{\,\,\,\,}r@{\,\,\,\,}r@{\,\,\,\,}|
@{\,\,\,\,}r@{\,\,\,\,}r@{\,\,\,\,}r@{\,\,\,\,}r@{\,\,\,\,}}
\hline
	 \multicolumn{2}{c}{}&
   \multicolumn{7}{c}{\textbf{Class A}}&
   \multicolumn{7}{c}{\textbf{Class B}}&
   \multicolumn{7}{c}{\textbf{Class C}}&
   \multicolumn{7}{c}{\textbf{All classes}}\\
\hline
\textbf{Category}&\textbf{$n$}&
  \textbf{$\gamma_1$}&\textbf{$\gamma_2$}&\textbf{$\gamma_3$}&\textbf{$\eta_1$}&\textbf{$\eta_2$}&\textbf{$\eta_3$}&\textbf{$\#$}&
	$\gamma_1$&$\gamma_2$&$\gamma_3$&$\eta_1$&$\eta_2$&$\eta_3$&$\#$&
	$\gamma_1$&$\gamma_2$&$\gamma_3$&$\eta_1$&$\eta_2$&$\eta_3$&$\#$&
	$\gamma_1$&$\gamma_2$&$\gamma_3$&$\eta_1$&$\eta_2$&$\eta_3$&$\#$\\	\hline
1&20&6.3&65.9&0.6&4&0&8&0&7.7&78.4&3.0&4&0&6&0&14.0&85.8&0.8&4&0&7&0&9.3&76.7&1.5&12&0&21&0\\
1&40&5.0&59.1&0.0&5&0&9&1&8.8&68.9&0.0&3&0&9&0&24.3&85.2&0.0&2&0&10&0&12.7&71.1&0.0&10&0&28&1\\
1&60&5.3&55.4&0.0&5&0&9&1&13.6&60.8&0.0&2&0&9&1&21.6&81.6&0.0&3&0&9&1&13.5&66.0&0.0&10&0&27&3\\
1&80&9.3&52.5&0.0&2&0&9&1&14.6&57.6&0.0&2&0&8&2&28.0&83.5&0.0&3&0&8&2&17.3&64.5&0.0&7&0&25&5\\
1&100&1.7&50.9&0.0&8&0&4&6&3.6&59.7&0.0&6&0&5&5&16.0&82.8&0.0&3&0&7&3&7.1&64.5&0.0&17&0&16&14\\
\hline
\multicolumn{2}{l}{\textbf{Category 1}}&\textbf{5.5}&\textbf{56.8}&\textbf{0.2}&\textbf{24}&\textbf{0}&\textbf{39}&\textbf{9}&\textbf{9.7}&\textbf{65.1}&\textbf{0.7}&\textbf{17}&\textbf{0}&\textbf{37}&\textbf{8}&\textbf{20.8}&\textbf{83.8}&\textbf{0.2}&\textbf{15}&\textbf{0}&\textbf{41}&\textbf{6}&\textbf{12.0}&\textbf{68.5}&\textbf{0.4}&\textbf{56}&\textbf{0}&\textbf{117}&\textbf{23}\\
\hline
2&20&0.0&0.0&0.0&10&10&10&0&0.0&0.0&0.0&10&10&10&0&0.0&0.0&0.0&10&10&10&0&0.0&0.0&0.0&30&30&30&0\\
2&40&0.0&84.3&41.5&10&1&2&0&0.0&89.8&46.4&10&1&1&0&0.0&90.0&55.4&10&1&1&0&0.0&88.1&47.7&30&3&4&0\\
2&60&0.0&87.3&29.1&10&0&0&0&0.0&98.6&33.3&10&0&0&0&0.0&100.0&60.6&10&0&0&0&0.0&95.3&41.0&30&0&0&0\\
2&80&0.0&82.4&11.8&10&0&0&0&0.0&97.0&21.3&10&0&0&0&0.0&100.0&44.0&10&0&0&0&0.0&93.1&25.7&30&0&0&0\\
2&100&0.0&87.0&34.2&10&0&0&0&0.0&99.8&50.0&10&0&0&0&0.0&100.0&75.3&10&0&0&0&0.0&95.6&53.2&30&0&0&0\\
\hline
\multicolumn{2}{l}{\textbf{Category 2}}&\textbf{0.0}&\textbf{68.2}&\textbf{23.3}&\textbf{50}&\textbf{11}&\textbf{12}&\textbf{0}&\textbf{0.0}&\textbf{77.1}&\textbf{30.2}&\textbf{50}&\textbf{11}&\textbf{11}&\textbf{0}&\textbf{0.0}&\textbf{78.0}&\textbf{47.0}&\textbf{50}&\textbf{11}&\textbf{11}&\textbf{0}&\textbf{0.0}&\textbf{74.4}&\textbf{33.5}&\textbf{150}&\textbf{33}&\textbf{34}&\textbf{0}\\
\hline
3&20&4.7&75.1&7.4&5&0&7&0&7.0&89.7&3.5&7&0&6&0&20.9&94.9&11.2&4&0&5&0&10.9&86.5&7.4&16&0&18&0\\
3&40&8.2&63.9&0.9&5&0&9&0&12.1&79.8&0.4&5&0&8&0&9.5&89.1&7.7&6&0&6&0&9.9&77.6&3.0&16&0&23&0\\
3&60&8.1&60.5&0.0&3&0&9&1&11.2&69.4&0.1&4&0&9&0&23.7&84.7&0.0&2&0&9&1&14.3&71.5&0.1&9&0&27&2\\
3&80&4.2&58.6&0.0&5&0&7&3&4.2&65.3&0.0&7&0&4&6&13.3&78.3&0.0&6&0&7&3&7.2&67.4&0.0&18&0&18&12\\
3&100&3.1&53.5&0.0&6&0&4&6&5.1&63.1&0.0&6&0&5&5&13.7&90.2&0.0&6&0&5&5&7.3&68.9&0.0&18&0&14&16\\
\hline
\multicolumn{2}{l}{\textbf{Category 3}}&\textbf{5.7}&\textbf{62.3}&\textbf{2.1}&\textbf{24}&\textbf{0}&\textbf{36}&\textbf{10}&\textbf{7.9}&\textbf{73.5}&\textbf{1.0}&\textbf{29}&\textbf{0}&\textbf{32}&\textbf{11}&\textbf{16.2}&\textbf{87.4}&\textbf{4.6}&\textbf{24}&\textbf{0}&\textbf{32}&\textbf{9}&\textbf{9.9}&\textbf{74.4}&\textbf{2.6}&\textbf{77}&\textbf{0}&\textbf{100}&\textbf{30}\\
\hline
4&20&0.0&0.0&0.0&10&10&10&0&0.0&0.0&0.0&10&10&10&0&0.0&0.0&0.0&10&10&10&0&0.0&0.0&0.0&30&30&30&0\\
4&40&0.0&84.4&50.3&10&1&3&0&0.0&89.7&51.6&10&1&2&0&0.0&90.0&65.3&10&1&1&0&0.0&88.0&55.8&30&3&6&0\\
4&60&0.0&84.2&20.5&10&0&0&0&0.0&96.4&23.7&10&0&0&0&0.0&100.0&48.1&10&0&1&0&0.0&93.5&30.8&30&0&1&0\\
4&80&0.0&82.1&17.0&10&0&0&0&0.0&97.3&22.6&10&0&0&0&0.0&100.0&57.9&10&0&0&0&0.0&93.1&32.5&30&0&0&0\\
4&100&0.0&84.9&25.7&10&0&0&0&0.0&98.5&45.4&10&0&0&0&0.0&100.0&72.1&10&0&0&0&0.0&94.5&47.7&30&0&0&0\\
\hline
\multicolumn{2}{l}{\textbf{Category 4}}&\textbf{0.0}&\textbf{67.1}&\textbf{22.7}&\textbf{50}&\textbf{11}&\textbf{13}&\textbf{0}&\textbf{0.0}&\textbf{76.4}&\textbf{28.7}&\textbf{50}&\textbf{11}&\textbf{12}&\textbf{0}&\textbf{0.0}&\textbf{78.0}&\textbf{48.7}&\textbf{50}&\textbf{11}&\textbf{12}&\textbf{0}&\textbf{0.0}&\textbf{73.8}&\textbf{33.4}&\textbf{150}&\textbf{33}&\textbf{37}&\textbf{0}\\
\hline
5&20&6.1&69.7&1.9&5&0&7&0&16.4&85.1&7.8&5&0&6&0&20.9&92.1&7.7&5&0&6&0&14.5&82.3&5.8&15&0&19&0\\
5&40&7.4&58.1&0.0&4&0&9&1&7.0&71.6&0.2&5&0&8&1&23.9&91.4&3.4&2&0&9&0&12.8&73.7&1.3&11&0&26&2\\
5&60&4.9&62.4&0.0&5&0&7&3&1.4&65.3&0.0&8&0&7&3&20.1&82.0&0.0&5&0&10&0&8.8&69.9&0.0&18&0&24&6\\
5&80&5.7&54.1&0.0&5&0&6&4&8.7&62.6&0.0&6&0&5&5&19.7&75.9&0.0&3&0&8&2&11.4&64.2&0.0&14&0&19&11\\
5&100&1.7&56.0&0.0&7&0&3&7&0.0&55.0&0.0&10&0&1&9&18.6&76.9&0.0&3&0&8&2&6.8&62.6&0.0&20&0&12&18\\
\hline
\multicolumn{2}{l}{\textbf{Category 5}}&\textbf{5.2}&\textbf{60.1}&\textbf{0.5}&\textbf{26}&\textbf{0}&\textbf{32}&\textbf{15}&\textbf{6.7}&\textbf{67.9}&\textbf{2.5}&\textbf{34}&\textbf{0}&\textbf{27}&\textbf{18}&\textbf{20.6}&\textbf{83.7}&\textbf{2.4}&\textbf{18}&\textbf{0}&\textbf{41}&\textbf{4}&\textbf{10.8}&\textbf{70.5}&\textbf{1.9}&\textbf{78}&\textbf{0}&\textbf{100}&\textbf{37}\\
\hline
6&20&0.0&0.0&0.0&10&10&10&0&0.0&0.0&0.0&10&10&10&0&0.0&0.0&0.0&10&10&10&0&0.0&0.0&0.0&30&30&30&0\\
6&40&0.0&49.2&20.8&10&5&5&0&0.0&50.0&21.3&10&5&6&0&0.0&50.0&29.7&10&5&5&0&0.0&49.7&23.9&30&15&16&0\\
6&60&0.0&87.2&8.4&10&0&0&0&0.0&98.9&17.3&10&0&0&0&0.0&100.0&46.2&10&0&0&0&0.0&95.4&24.0&30&0&0&0\\
6&80&0.0&93.9&42.4&10&0&0&0&0.0&100.0&60.0&10&0&0&0&0.0&100.0&83.8&10&0&0&0&0.0&98.0&62.1&30&0&0&0\\
6&100&0.0&87.3&23.8&10&0&0&0&0.0&99.4&38.7&10&0&0&0&0.0&100.0&73.2&10&0&0&0&0.0&95.6&45.2&30&0&0&0\\
\hline
\multicolumn{2}{l}{\textbf{Category 6}}&\textbf{0.0}&\textbf{63.5}&\textbf{19.1}&\textbf{50}&\textbf{15}&\textbf{15}&\textbf{0}&\textbf{0.0}&\textbf{69.7}&\textbf{27.5}&\textbf{50}&\textbf{15}&\textbf{16}&\textbf{0}&\textbf{0.0}&\textbf{70.0}&\textbf{46.6}&\textbf{50}&\textbf{15}&\textbf{15}&\textbf{0}&\textbf{0.0}&\textbf{67.7}&\textbf{31.0}&\textbf{150}&\textbf{45}&\textbf{46}&\textbf{0}\\
\hline
7&20&11.8&71.4&10.2&2&0&1&0&21.5&84.1&18.1&1&0&1&0&42.4&96.0&28.1&1&0&2&0&25.2&83.9&18.8&4&0&4&0\\
7&40&2.3&64.0&0.7&5&0&5&1&5.6&73.2&1.2&6&0&7&1&24.4&96.0&1.4&1&0&9&0&10.8&77.7&1.1&12&0&21&2\\
7&60&0.0&59.9&0.4&10&0&0&9&0.0&74.0&8.3&10&0&0&8&4.1&88.2&5.3&8&0&5&2&1.4&74.1&5.4&28&0&5&19\\
7&80&0.0&57.8&0.0&10&0&1&9&0.0&63.6&&10&0&&10&7.4&79.2&0.0&7&0&3&7&2.5&66.9&0.0&27&0&4&26\\
7&100&0.0&56.1&&10&0&&10&0.0&59.9&&10&0&&10&2.8&90.1&8.4&9&0&1&8&0.9&68.7&8.4&29&0&1&28\\
\hline
\multicolumn{2}{l}{\textbf{Category 7}}&\textbf{2.8}&\textbf{61.8}&\textbf{5.2}&\textbf{37}&\textbf{0}&\textbf{7}&\textbf{29}&\textbf{5.4}&\textbf{70.9}&\textbf{10.0}&\textbf{37}&\textbf{0}&\textbf{8}&\textbf{29}&\textbf{16.2}&\textbf{89.9}&\textbf{10.7}&\textbf{26}&\textbf{0}&\textbf{20}&\textbf{17}&\textbf{8.1}&\textbf{74.2}&\textbf{8.9}&\textbf{100}&\textbf{0}&\textbf{35}&\textbf{75}\\
\hline
8&20&17.6&71.0&15.6&1&0&1&0&20.2&82.0&20.6&1&0&1&0&47.7&95.9&28.4&0&0&1&0&28.5&83.0&21.5&2&0&3&0\\
8&40&0.0&59.2&1.0&10&0&7&2&5.9&75.3&0.0&4&0&8&2&12.6&93.0&0.0&5&0&10&0&6.2&75.8&0.3&19&0&25&4\\
8&60&0.0&64.4&4.2&10&0&1&6&0.7&74.5&0.0&9&0&3&7&5.4&90.0&16.7&7&0&5&4&2.1&76.3&9.0&26&0&9&17\\
8&80&0.0&55.6&&10&0&&10&0.0&57.6&&10&0&&10&1.3&93.9&0.0&9&0&2&8&0.4&69.1&0.0&29&0&2&28\\
8&100&0.0&55.8&5.1&10&0&0&9&0.0&63.5&0.0&10&0&1&9&0.5&92.1&0.0&9&0&1&9&0.2&70.5&1.7&29&0&2&27\\
\hline
\multicolumn{2}{l}{\textbf{Category 8}}&\textbf{3.5}&\textbf{61.2}&\textbf{8.1}&\textbf{41}&\textbf{0}&\textbf{9}&\textbf{27}&\textbf{5.3}&\textbf{70.6}&\textbf{9.4}&\textbf{34}&\textbf{0}&\textbf{13}&\textbf{28}&\textbf{13.5}&\textbf{93.0}&\textbf{13.2}&\textbf{30}&\textbf{0}&\textbf{19}&\textbf{21}&\textbf{7.5}&\textbf{74.9}&\textbf{10.5}&\textbf{105}&\textbf{0}&\textbf{41}&\textbf{76}\\
\hline
9&20&0.0&61.4&0.0&10&0&10&0&0.0&71.4&0.0&10&0&10&0&0.0&81.1&0.0&10&0&10&0&0.0&71.3&0.0&30&0&30&0\\
9&40&2.5&57.5&0.0&7&0&10&0&2.9&75.8&0.0&7&0&10&0&4.7&80.2&0.0&9&0&10&0&3.4&71.1&0.0&23&0&30&0\\
9&60&1.6&58.1&0.0&7&0&10&0&2.6&66.1&0.0&7&0&10&0&4.6&81.0&0.0&8&0&10&0&2.9&68.4&0.0&22&0&30&0\\
9&80&2.1&56.2&0.0&5&0&10&0&2.7&65.3&0.0&6&0&10&0&15.2&79.9&0.0&6&0&10&0&6.7&67.1&0.0&17&0&30&0\\
9&100&1.7&55.1&0.0&5&0&10&0&2.9&65.0&0.0&5&0&10&0&5.2&77.9&0.0&5&0&10&0&3.3&66.0&0.0&15&0&30&0\\
\hline
\multicolumn{2}{l}{\textbf{Category 9}}&\textbf{1.6}&\textbf{57.7}&\textbf{0.0}&\textbf{34}&\textbf{0}&\textbf{50}&\textbf{0}&\textbf{2.2}&\textbf{68.7}&\textbf{0.0}&\textbf{35}&\textbf{0}&\textbf{50}&\textbf{0}&\textbf{5.9}&\textbf{80.0}&\textbf{0.0}&\textbf{38}&\textbf{0}&\textbf{50}&\textbf{0}&\textbf{3.2}&\textbf{68.8}&\textbf{0.0}&\textbf{107}&\textbf{0}&\textbf{150}&\textbf{0}\\
\hline
10&20&9.4&72.5&7.8&5&0&6&0&15.0&80.1&13.9&4&0&4&0&13.0&96.5&3.4&6&0&7&0&12.5&83.1&8.4&15&0&17&0\\
10&40&3.0&68.5&5.0&8&0&5&0&2.6&75.9&1.4&7&0&7&0&8.6&91.8&13.2&7&0&6&0&4.8&78.7&6.5&22&0&18&0\\
10&60&0.0&64.9&4.6&10&0&4&1&5.7&76.0&6.9&6&0&4&1&15.3&98.7&6.5&6&0&5&0&7.0&79.9&6.0&22&0&13&2\\
10&80&0.0&60.8&3.4&10&0&4&3&0.7&68.1&2.6&8&0&5&1&10.3&98.0&4.1&7&0&6&1&3.7&75.7&3.3&25&0&15&5\\
10&100&0.4&56.8&3.1&9&0&3&3&0.6&66.7&4.7&8&0&3&4&3.2&92.0&10.8&9&0&4&1&1.4&71.8&6.7&26&0&10&8\\
\hline
\multicolumn{2}{l}{\textbf{Category 10}}&\textbf{2.6}&\textbf{64.7}&\textbf{5.0}&\textbf{42}&\textbf{0}&\textbf{22}&\textbf{7}&\textbf{5.0}&\textbf{73.4}&\textbf{6.1}&\textbf{33}&\textbf{0}&\textbf{23}&\textbf{6}&\textbf{10.1}&\textbf{95.4}&\textbf{7.6}&\textbf{35}&\textbf{0}&\textbf{28}&\textbf{2}&\textbf{5.9}&\textbf{77.8}&\textbf{6.3}&\textbf{110}&\textbf{0}&\textbf{73}&\textbf{15}\\
\hline
\hline
\multicolumn{2}{l}{\textbf{All categories}}&\textbf{2.7}&\textbf{62.3}&\textbf{9.6}&\textbf{378}&\textbf{37}&\textbf{235}&\textbf{97}&\textbf{4.2}&\textbf{71.3}&\textbf{12.9}&\textbf{369}&\textbf{37}&\textbf{229}&\textbf{100}&\textbf{10.3}&\textbf{83.9}&\textbf{19.3}&\textbf{336}&\textbf{37}&\textbf{269}&\textbf{59}&\textbf{5.7}&\textbf{72.5}&\textbf{14.1}&\textbf{1083}&\textbf{111}&\textbf{733}&\textbf{256}\\
\hline
\end{tabular}
}
\end{tiny}
\end{sidewaystable}

\begin{sidewaystable}
 \caption{Statistics of the runtime of {\PH}}\label{Tab:LBRT}
\vskip+0.5cm
\begin{tiny}{
\hskip-0.8cm
\hspace*{1cm}
\begin{tabular}{@{\,\,\,\,}r@{\,\,\,\,}r@{\,\,\,\,}||
@{\,\,\,\,}r@{\,\,\,\,}r@{\,\,\,\,}r@{\,\,\,\,}r@{\,\,\,\,}r@{\,\,\,\,}||
@{\,\,\,\,}r@{\,\,\,\,}r@{\,\,\,\,}r@{\,\,\,\,}r@{\,\,\,\,}r@{\,\,\,\,}||
@{\,\,\,\,}r@{\,\,\,\,}r@{\,\,\,\,}r@{\,\,\,\,}r@{\,\,\,\,}r@{\,\,\,\,}||
@{\,\,\,\,}r@{\,\,\,\,}r@{\,\,\,\,}r@{\,\,\,\,}r@{\,\,\,\,}r@{\,\,\,\,}}
\hline
   \multicolumn{2}{c}{}&
	 \multicolumn{5}{c}{\textbf{Class A}}&
   \multicolumn{5}{c}{\textbf{Class B}}&
   \multicolumn{5}{c}{\textbf{Class C}}&
   \multicolumn{5}{c}{\textbf{All classes}}\\
\hline	
\textbf{Category}&	$n$	&$RT$&$Q_2$&$\underline{RT}$&$\overbar{RT}$&$\#$
        &$RT$&$Q_2$&$\underline{RT}$&$\overbar{RT}$&$\#$
        &$RT$&$Q_2$&$\underline{RT}$&$\overbar{RT}$&$\#$
        &$RT$&$Q_2$&$\underline{RT}$&$\overbar{RT}$&$\#$\\  \hline
1&20&2.14&1.86&0.22&4.74&0&2.04&1.97&0.29&5.01&0&1.99&1.90&0.27&4.11&0&2.05&1.90&0.22&5.01&0\\
1&40&8.87&4.92&0.65&24.24&1&32.87&5.65&0.62&255.06&0&10.11&4.32&0.47&39.24&0&17.57&4.61&0.47&255.06&1\\
1&60&53.65&16.54&4.88&313.27&1&51.58&12.94&3.86&239.62&1&79.90&14.31&6.81&452.46&1&61.71&16.19&3.86&452.46&3\\
1&80&124.08&51.46&15.52&450.12&1&539.04&41.87&15.25&3587.90&2&32.16&25.35&13.46&63.91&2&227.45&33.14&13.46&3587.90&5\\
1&100&609.61&532.60&95.66&1277.58&6&316.59&281.97&69.87&641.00&5&332.34&257.04&50.84&1129.49&3&396.74&300.79&50.84&1277.58&14\\
\hline
\multicolumn{2}{l}{\textbf{Category 1}}&\textbf{100.96}&\textbf{15.52}&\textbf{0.22}&\textbf{1277.58}&\textbf{9}&\textbf{159.73}&\textbf{12.46}&\textbf{0.29}&\textbf{3587.90}&\textbf{8}&\textbf{77.81}&\textbf{12.42}&\textbf{0.27}&\textbf{1129.49}&\textbf{6}&\textbf{112.37}&\textbf{13.46}&\textbf{0.22}&\textbf{3587.90}&\textbf{23}\\
\hline
2&20&0.01&0.01&0.01&0.01&0&0.01&0.01&0.01&0.01&0&0.01&0.01&0.01&0.02&0&0.01&0.01&0.01&0.02&0\\
2&40&0.92&0.50&0.01&2.95&0&0.82&0.61&0.01&1.94&0&0.74&0.10&0.01&2.32&0&0.82&0.10&0.01&2.95&0\\
2&60&0.24&0.23&0.22&0.29&0&0.22&0.22&0.19&0.25&0&0.42&0.20&0.15&2.43&0&0.29&0.23&0.15&2.43&0\\
2&80&1.33&0.64&0.44&5.31&0&1.33&0.70&0.34&4.49&0&1.79&1.61&0.39&4.30&0&1.48&1.02&0.34&5.31&0\\
2&100&1.49&0.98&0.53&3.63&0&1.78&1.00&0.72&4.54&0&0.94&0.71&0.51&2.51&0&1.40&0.93&0.51&4.54&0\\
\hline
\multicolumn{2}{l}{\textbf{Category 2}}&\textbf{0.80}&\textbf{0.36}&\textbf{0.01}&\textbf{5.31}&\textbf{0}&\textbf{0.83}&\textbf{0.30}&\textbf{0.01}&\textbf{4.54}&\textbf{0}&\textbf{0.78}&\textbf{0.31}&\textbf{0.01}&\textbf{4.30}&\textbf{0}&\textbf{0.80}&\textbf{0.32}&\textbf{0.01}&\textbf{5.31}&\textbf{0}\\
\hline
3&20&2.01&2.52&0.23&3.15&0&1.41&1.77&0.19&2.46&0&1.41&1.46&0.16&3.75&0&1.61&1.85&0.16&3.75&0\\
3&40&23.92&6.84&0.54&124.81&0&38.97&5.05&0.63&304.06&0&216.39&4.49&0.56&2127.61&0&93.09&5.50&0.54&2127.61&0\\
3&60&39.66&26.57&6.62&126.54&1&108.69&18.63&4.78&704.02&0&18.40&12.77&6.91&42.99&1&57.48&19.92&4.78&704.02&2\\
3&80&462.72&199.61&40.84&1750.29&3&100.84&103.68&27.72&168.29&6&58.55&34.68&18.30&209.24&3&225.13&72.54&18.30&1750.29&12\\
3&100&270.91&218.31&50.99&596.04&6&525.85&87.81&31.07&2352.10&5&54.94&54.79&27.86&79.38&5&284.83&73.95&27.86&2352.10&16\\
\hline
\multicolumn{2}{l}{\textbf{Category 3}}&\textbf{123.47}&\textbf{17.31}&\textbf{0.23}&\textbf{1750.29}&\textbf{10}&\textbf{115.98}&\textbf{12.78}&\textbf{0.19}&\textbf{2352.10}&\textbf{11}&\textbf{73.86}&\textbf{8.30}&\textbf{0.16}&\textbf{2127.61}&\textbf{9}&\textbf{104.09}&\textbf{10.72}&\textbf{0.16}&\textbf{2352.10}&\textbf{30}\\
\hline
4&20&0.01&0.01&0.01&0.01&0&0.01&0.01&0.01&0.01&0&0.01&0.01&0.01&0.01&0&0.01&0.01&0.01&0.01&0\\
4&40&0.44&0.02&0.01&1.46&0&0.38&0.02&0.01&1.66&0&0.38&0.02&0.01&1.53&0&0.40&0.02&0.01&1.66&0\\
4&60&0.24&0.23&0.21&0.28&0&0.23&0.23&0.20&0.27&0&0.21&0.21&0.15&0.25&0&0.22&0.23&0.15&0.28&0\\
4&80&0.84&0.53&0.39&1.44&0&1.44&1.60&0.37&2.76&0&1.65&1.57&0.35&4.09&0&1.31&1.41&0.35&4.09&0\\
4&100&1.22&1.01&0.74&3.56&0&2.71&2.06&0.51&6.26&0&1.93&0.89&0.48&5.00&0&1.95&0.98&0.48&6.26&0\\
\hline
\multicolumn{2}{l}{\textbf{Category 4}}&\textbf{0.55}&\textbf{0.25}&\textbf{0.01}&\textbf{3.56}&\textbf{0}&\textbf{0.95}&\textbf{0.25}&\textbf{0.01}&\textbf{6.26}&\textbf{0}&\textbf{0.83}&\textbf{0.24}&\textbf{0.01}&\textbf{5.00}&\textbf{0}&\textbf{0.78}&\textbf{0.25}&\textbf{0.01}&\textbf{6.26}&\textbf{0}\\
\hline
5&20&2.17&2.49&0.18&5.36&0&2.13&1.87&0.18&5.39&0&1.34&0.74&0.21&3.85&0&1.88&1.87&0.18&5.39&0\\
5&40&21.59&5.59&0.82&66.19&1&64.12&5.18&2.38&327.80&1&5.05&4.07&0.54&12.64&0&29.35&5.20&0.54&327.80&2\\
5&60&170.16&19.43&10.73&643.32&3&39.78&31.33&4.09&93.73&3&47.54&12.91&6.53&293.40&0&81.04&17.73&4.09&643.32&6\\
5&80&832.08&526.51&26.06&2395.61&4&138.64&101.81&25.94&359.81&5&209.72&64.21&16.76&892.81&2&387.55&101.81&16.76&2395.61&11\\
5&100&366.41&203.19&184.46&711.58&7&1436.20&1436.20&1436.20&1436.20&9&376.63&143.92&34.87&1883.37&2&462.37&193.83&34.87&1883.37&18\\
\hline
\multicolumn{2}{l}{\textbf{Category 5}}&\textbf{214.25}&\textbf{12.04}&\textbf{0.18}&\textbf{2395.61}&\textbf{15}&\textbf{93.94}&\textbf{6.53}&\textbf{0.18}&\textbf{1436.20}&\textbf{18}&\textbf{113.70}&\textbf{10.74}&\textbf{0.21}&\textbf{1883.37}&\textbf{4}&\textbf{139.25}&\textbf{9.98}&\textbf{0.18}&\textbf{2395.61}&\textbf{37}\\
\hline
6&20&0.01&0.01&0.01&0.01&0&0.01&0.01&0.01&0.01&0&0.01&0.01&0.01&0.02&0&0.01&0.01&0.01&0.02&0\\
6&40&0.69&0.02&0.01&2.78&0&0.37&0.02&0.01&1.26&0&0.24&0.03&0.01&1.05&0&0.43&0.02&0.01&2.78&0\\
6&60&0.22&0.22&0.19&0.24&0&0.21&0.21&0.15&0.25&0&0.19&0.20&0.15&0.24&0&0.21&0.21&0.15&0.25&0\\
6&80&0.77&0.42&0.34&2.06&0&1.26&0.90&0.29&3.33&0&0.58&0.32&0.26&2.87&0&0.87&0.40&0.26&3.33&0\\
6&100&0.73&0.70&0.65&0.85&0&2.74&3.68&0.59&4.57&0&0.94&0.56&0.47&3.98&0&1.47&0.70&0.47&4.57&0\\
\hline
\multicolumn{2}{l}{\textbf{Category 6}}&\textbf{0.48}&\textbf{0.23}&\textbf{0.01}&\textbf{2.78}&\textbf{0}&\textbf{0.91}&\textbf{0.22}&\textbf{0.01}&\textbf{4.57}&\textbf{0}&\textbf{0.39}&\textbf{0.21}&\textbf{0.01}&\textbf{3.98}&\textbf{0}&\textbf{0.60}&\textbf{0.23}&\textbf{0.01}&\textbf{4.57}&\textbf{0}\\
\hline
7&20&2.81&2.57&1.70&4.16&0&3.21&3.03&0.78&7.26&0&3.34&3.28&1.12&5.85&0&3.12&2.89&0.78&7.26&0\\
7&40&588.34&15.86&4.14&1995.20&1&492.04&11.88&3.42&3159.14&1&10.67&5.98&3.11&32.16&0&351.08&12.07&3.11&3159.14&2\\
7&60&23.84&23.84&23.84&23.84&9&330.31&330.31&65.48&595.14&8&293.33&28.11&5.76&2008.52&2&275.55&39.31&5.76&2008.52&19\\
7&80&119.79&119.79&119.79&119.79&9&&&&&10&897.89&30.61&29.59&2633.47&7&703.36&75.20&29.59&2633.47&26\\
7&100&&&&&10&&&&&10&2976.24&2976.24&2523.98&3428.51&8&2976.24&2976.24&2523.98&3428.51&28\\
\hline
\multicolumn{2}{l}{\textbf{Category 7}}&\textbf{260.32}&\textbf{4.16}&\textbf{1.70}&\textbf{1995.20}&\textbf{29}&\textbf{243.86}&\textbf{6.87}&\textbf{0.78}&\textbf{3159.14}&\textbf{29}&\textbf{337.36}&\textbf{6.33}&\textbf{1.12}&\textbf{3428.51}&\textbf{17}&\textbf{289.61}&\textbf{6.33}&\textbf{0.78}&\textbf{3428.51}&\textbf{75}\\
\hline
8&20&7.24&3.82&2.70&32.54&0&4.04&3.17&1.39&7.20&0&3.60&3.59&1.80&5.46&0&4.96&3.50&1.39&32.54&0\\
8&40&122.28&35.51&15.01&530.00&2&154.01&28.13&7.33&772.32&2&11.00&7.32&4.16&41.59&0&89.24&18.35&4.16&772.32&4\\
8&60&245.57&89.54&35.19&768.01&6&790.76&28.03&22.36&2321.90&7&109.60&49.69&9.82&412.88&4&308.63&47.38&9.82&2321.90&17\\
8&80&&&&&10&&&&&10&2917.53&2917.53&2397.57&3437.50&8&2917.53&2917.53&2397.57&3437.50&28\\
8&100&144.22&144.22&144.22&144.22&9&1479.94&1479.94&1479.94&1479.94&9&311.61&311.61&311.61&311.61&9&645.26&311.61&144.22&1479.94&27\\
\hline
\multicolumn{2}{l}{\textbf{Category 8}}&\textbf{94.66}&\textbf{18.78}&\textbf{2.70}&\textbf{768.01}&\textbf{27}&\textbf{232.94}&\textbf{10.91}&\textbf{1.39}&\textbf{2321.90}&\textbf{28}&\textbf{239.66}&\textbf{7.00}&\textbf{1.80}&\textbf{3437.50}&\textbf{21}&\textbf{192.60}&\textbf{9.67}&\textbf{1.39}&\textbf{3437.50}&\textbf{76}\\
\hline
9&20&0.63&0.23&0.21&3.00&0&0.51&0.27&0.20&1.70&0&0.55&0.24&0.18&1.68&0&0.57&0.24&0.18&3.00&0\\
9&40&3.69&2.49&1.03&12.64&0&3.04&1.99&0.86&11.66&0&1.48&1.24&0.65&2.61&0&2.74&1.79&0.65&12.64&0\\
9&60&15.33&13.71&5.64&37.08&0&15.28&12.44&6.30&32.53&0&6.94&5.68&3.81&13.84&0&12.52&10.98&3.81&37.08&0\\
9&80&58.46&51.81&15.29&139.71&0&44.20&42.25&19.79&71.89&0&19.56&18.59&8.25&33.52&0&40.74&32.51&8.25&139.71&0\\
9&100&198.30&113.16&0.00&592.43&0&146.86&125.91&75.34&305.05&0&54.46&42.40&26.82&110.12&0&133.21&94.07&0.00&592.43&0\\
\hline
\multicolumn{2}{l}{\textbf{Category 9}}&\textbf{55.28}&\textbf{12.52}&\textbf{0.00}&\textbf{592.43}&\textbf{0}&\textbf{41.98}&\textbf{12.56}&\textbf{0.20}&\textbf{305.05}&\textbf{0}&\textbf{16.60}&\textbf{5.68}&\textbf{0.18}&\textbf{110.12}&\textbf{0}&\textbf{37.95}&\textbf{10.98}&\textbf{0.00}&\textbf{592.43}&\textbf{0}\\
\hline
10&20&2.03&1.83&0.20&4.29&0&1.75&1.85&0.53&3.08&0&1.73&1.74&0.24&3.40&0&1.84&1.76&0.20&4.29&0\\
10&40&21.96&2.22&0.60&175.73&0&13.09&3.38&0.43&103.75&0&2.27&1.65&0.36&5.26&0&12.44&2.31&0.36&175.73&0\\
10&60&14.80&10.46&6.06&35.88&1&15.51&8.66&5.08&65.43&1&8.28&6.74&0.84&34.47&0&12.70&8.39&0.84&65.43&2\\
10&80&51.53&25.45&10.41&222.78&3&220.03&20.10&13.77&1812.68&1&25.27&18.90&4.69&57.86&1&102.74&20.10&4.69&1812.68&5\\
10&100&145.47&38.22&15.89&801.33&3&616.26&61.51&27.76&3424.50&4&70.61&26.18&11.65&424.24&1&243.24&36.50&11.65&3424.50&8\\
\hline
\multicolumn{2}{l}{\textbf{Category 10}}&\textbf{40.75}&\textbf{10.03}&\textbf{0.20}&\textbf{801.33}&\textbf{7}&\textbf{135.59}&\textbf{8.39}&\textbf{0.43}&\textbf{3424.50}&\textbf{6}&\textbf{20.53}&\textbf{5.02}&\textbf{0.24}&\textbf{424.24}&\textbf{2}&\textbf{64.47}&\textbf{7.29}&\textbf{0.20}&\textbf{3424.50}&\textbf{15}\\
\hline
\hline
\multicolumn{2}{l}{\textbf{All categories}}&\textbf{71.54}&\textbf{2.57}&\textbf{0.00}&\textbf{2395.61}&\textbf{97}&\textbf{81.71}&\textbf{2.58}&\textbf{0.01}&\textbf{3587.90}&\textbf{100}&\textbf{71.84}&\textbf{2.87}&\textbf{0.01}&\textbf{3437.49}&\textbf{59}&\textbf{74.91}&\textbf{2.65}&\textbf{0.00}&\textbf{3587.90}&\textbf{256}\\
\hline
\end{tabular}
}
\end{tiny}
\end{sidewaystable}

%>>>>>>>>>>>>>>>>>>>>>>>>>>>>>>>>>>>>>>>>>>>>>>>>>>>>>>>>>>>>>>>>>>>>>>>>>>>>>>>>>>>>>>>>>>>>>>>>>>>>>>>>>>>>>>>>>
\subsection{Quality of the New Upper Bounds}\label{ssec:CEUB}

Table \ref{Tab:UB} displays for each class, category, and problem size, the average percent deviation $\delta^{F},\ \delta^{A},\ \delta^{E},$ of the upper bounds $L_{max}^{F},\ L_{max}^{A},\ L_{max}^{E},$ obtained respectively by {\FG}, {\Alga} and {\MIP}, from the best known lower bound $LB^*.$  In addition, Table \ref{Tab:UB} displays $\eta^{\bullet},$ the number of times $L_{max}^{\bullet}=LB^*,\ \bullet=F,\ A,\ E$.

\begin{sidewaystable}
 \caption{Comparison of upper bounds}\label{Tab:UB}
\begin{tiny}{
\hspace*{1cm}
 \begin{tabular}{r@{\,\,\,\,}r@{\,\,\,\,}||
@{\,\,\,\,}r@{\,\,\,\,}r@{\,\,\,\,}r@{\,\,\,\,}|
@{\,\,\,\,}r@{\,\,\,\,}r@{\,\,\,\,}r@{\,\,\,\,}||
@{\,\,\,\,}r@{\,\,\,\,}r@{\,\,\,\,}r@{\,\,\,\,}|
@{\,\,\,\,}r@{\,\,\,\,}r@{\,\,\,\,}r@{\,\,\,\,}||
@{\,\,\,\,}r@{\,\,\,\,}r@{\,\,\,\,}r@{\,\,\,\,}|
@{\,\,\,\,}r@{\,\,\,\,}r@{\,\,\,\,}r@{\,\,\,\,}||
@{\,\,\,\,}r@{\,\,\,\,}r@{\,\,\,\,}r@{\,\,\,\,}|
@{\,\,\,\,}r@{\,\,\,\,}r@{\,\,\,\,}r@{\,\,\,\,}}
\hline
\multicolumn{2}{c}{} & \multicolumn{6}{c}{\textbf{Class A}}& \multicolumn{6}{c}{\textbf{Class B}} &\multicolumn{6}{c}{\textbf{Class C}}
&\multicolumn{6}{c}{\textbf{All classes}}\\
\hline
\textbf{Category}&	$n$	&
   	$\delta^{F}$	&	$\delta^{A}$	&	$\delta^{E}$	&	
   	$\eta^{F}$	  &	$\eta^{A}$	  &	$\eta^{E}$	  &	
   	$\delta^{F}$	&	$\delta^{A}$	&	$\delta^{E}$	&	
   	$\eta^{F}$	  &	$\eta^{A}$	  &	$\eta^{E}$	  &	
   	$\delta^{F}$	&	$\delta^{A}$	&	$\delta^{E}$	&	
   	$\eta^{F}$	  &	$\eta^{A}$	  &	$\eta^{E}$	  &	
   	$\delta^{F}$	&	$\delta^{A}$	&	$\delta^{E}$	&	
    $\eta^{F}$    &$\eta^{A}$     &$\eta^{E}$	    \\
\hline
1&20&18.0&3.6&0.0&4&7&10&21.7&1.4&0.0&3&8&10&51.6&4.9&0.0&1&6&10&30.5&3.3&0.0&8&21&30\\
1&40&24.3&8.4&8.9&1&2&4&27.6&9.4&13.2&1&4&5&45.3&18.7&11.6&1&4&6&32.4&12.2&11.2&3&10&15\\
1&60&22.1&9.0&50.0&0&3&0&36.5&11.2&58.4&0&2&0&53.9&33.1&132.1&0&4&1&37.5&17.8&78.4&0&9&1\\
1&80&19.7&5.6&147.3&1&4&0&34.8&11.8&346.7&1&3&0&69.8&23.4&683.5&0&6&0&41.4&13.6&382.5&2&13&0\\
1&100&21.9&10.0&523.9&0&1&0&39.6&17.8&913.9&1&1&0&104.0&58.1&2147.0&0&1&0&55.2&28.6&1194.9&1&3&0\\
\hline
\multicolumn{2}{l}{\textbf{Category 1}}&\textbf{21.2}&\textbf{7.3}&\textbf{146.0}&\textbf{6}&\textbf{17}&\textbf{14}&\textbf{32.0}&\textbf{10.3}&\textbf{266.4}&\textbf{6}&\textbf{18}&\textbf{15}&\textbf{64.9}&\textbf{27.6}&\textbf{602.6}&\textbf{2}&\textbf{21}&\textbf{17}&\textbf{39.4}&\textbf{15.1}&\textbf{334.8}&\textbf{14}&\textbf{56}&\textbf{46}\\
\hline
2&20&0.0&0.0&0.0&10&10&10&2.2&0.0&0.0&9&10&10&1.7&0.0&0.0&8&10&10&1.3&0.0&0.0&27&30&30\\
2&40&6.4&4.7&7.9&5&8&0&10.6&3.7&8.5&6&8&0&33.7&4.2&28.7&4&8&1&16.9&4.2&15.0&15&24&1\\
2&60&44.6&1.3&406.4&2&5&0&19.3&5.0&610.5&2&3&0&42.7&11.6&1102.5&3&5&0&35.5&6.0&709.8&7&13&0\\
2&80&11.0&4.8&4132.6&3&5&0&8.4&1.6&4755.3&3&5&0&22.3&11.0&11144.7&2&4&0&13.9&5.8&6743.8&8&14&0\\
2&100&1.7&0.9&5579.3&3&5&0&4.6&4.1&9018.8&1&1&0&14.9&9.2&20412.2&1&2&0&7.1&4.7&11368.7&5&8&0\\
\hline
\multicolumn{2}{l}{\textbf{Category 2}}&\textbf{12.8}&\textbf{2.3}&\textbf{2025.2}&\textbf{23}&\textbf{33}&\textbf{10}&\textbf{9.0}&\textbf{2.9}&\textbf{2886.8}&\textbf{21}&\textbf{27}&\textbf{10}&\textbf{23.1}&\textbf{7.2}&\textbf{6254.5}&\textbf{18}&\textbf{29}&\textbf{11}&\textbf{14.9}&\textbf{4.1}&\textbf{3716.3}&\textbf{62}&\textbf{89}&\textbf{31}\\
\hline
3&20&39.6&9.5&2.7&1&4&9&66.9&23.0&8.3&2&5&9&63.7&8.4&0.0&3&6&10&56.7&13.7&3.7&6&15&28\\
3&40&28.2&15.4&14.6&1&2&3&50.0&29.5&29.6&0&2&3&63.3&36.8&39.8&1&3&4&47.2&27.3&28.0&2&7&10\\
3&60&37.1&13.7&153.1&0&1&0&52.3&22.0&130.5&0&1&0&88.4&45.1&166.5&0&1&0&59.3&26.9&150.1&0&3&0\\
3&80&47.6&17.5&446.6&0&0&0&82.5&44.7&982.9&0&0&0&131.7&74.4&778.4&0&2&0&87.3&45.5&736.0&0&2&0\\
3&100&38.4&17.8&873.6&0&1&0&60.2&30.3&1457.3&0&2&0&144.9&89.5&3003.7&0&0&0&81.1&45.8&1778.2&0&3&0\\
\hline
\multicolumn{2}{l}{\textbf{Category 3}}&\textbf{38.2}&\textbf{14.8}&\textbf{298.1}&\textbf{2}&\textbf{8}&\textbf{12}&\textbf{62.4}&\textbf{29.9}&\textbf{521.7}&\textbf{2}&\textbf{10}&\textbf{12}&\textbf{98.4}&\textbf{50.8}&\textbf{797.7}&\textbf{4}&\textbf{12}&\textbf{14}&\textbf{66.3}&\textbf{31.8}&\textbf{539.2}&\textbf{8}&\textbf{30}&\textbf{38}\\
\hline
4&20&19.6&0.0&0.0&6&10&10&5.2&0.0&0.0&8&10&10&2.8&0.0&0.0&7&10&10&9.2&0.0&0.0&21&30&30\\
4&40&12.1&0.6&1.7&1&5&3&30.1&6.7&12.3&1&2&0&341.4&54.4&83.9&2&5&2&127.9&20.5&32.6&4&12&5\\
4&60&77.2&23.1&471.7&0&2&0&78.7&27.3&514.0&0&1&0&130.0&39.0&755.0&0&2&0&95.3&29.8&584.0&0&5&0\\
4&80&50.5&17.6&1829.2&0&0&0&54.0&23.1&1812.3&0&0&0&96.2&33.6&8226.6&0&0&0&66.9&24.8&3808.8&0&0&0\\
4&100&24.3&11.1&5890.4&0&0&0&36.5&14.3&8423.9&0&0&0&129.5&50.9&21683.5&1&1&0&63.4&25.4&11665.3&1&1&0\\
\hline
\multicolumn{2}{l}{\textbf{Category 4}}&\textbf{36.8}&\textbf{10.5}&\textbf{1662.4}&\textbf{7}&\textbf{17}&\textbf{13}&\textbf{40.9}&\textbf{14.3}&\textbf{2152.5}&\textbf{9}&\textbf{13}&\textbf{10}&\textbf{140.0}&\textbf{35.6}&\textbf{5782.9}&\textbf{10}&\textbf{18}&\textbf{12}&\textbf{72.5}&\textbf{20.1}&\textbf{3174.6}&\textbf{26}&\textbf{48}&\textbf{35}\\
\hline
5&20&16.8&5.4&3.0&2&6&7&27.9&1.6&0.0&2&8&10&95.0&4.5&0.0&4&9&10&46.6&3.8&1.0&8&23&27\\
5&40&27.5&11.3&14.6&0&3&4&34.1&13.9&16.7&2&4&4&63.5&21.8&11.1&0&3&5&41.7&15.7&14.2&2&10&13\\
5&60&32.5&14.5&50.7&0&2&0&53.2&31.0&84.5&0&3&0&63.9&32.5&132.0&1&2&0&49.9&26.0&88.9&1&7&0\\
5&80&35.1&14.0&328.4&0&2&0&63.1&28.0&440.8&0&2&0&70.0&30.0&777.2&0&2&0&56.1&24.0&515.5&0&6&0\\
5&100&40.7&18.0&630.6&0&1&0&66.4&40.9&1141.8&0&0&0&158.5&90.5&2220.4&0&1&0&88.5&49.8&1330.9&0&2&0\\
\hline
\multicolumn{2}{l}{\textbf{Category 5}}&\textbf{30.5}&\textbf{12.6}&\textbf{208.6}&\textbf{2}&\textbf{14}&\textbf{11}&\textbf{48.9}&\textbf{23.1}&\textbf{336.8}&\textbf{4}&\textbf{17}&\textbf{14}&\textbf{90.2}&\textbf{35.9}&\textbf{638.3}&\textbf{5}&\textbf{17}&\textbf{15}&\textbf{56.5}&\textbf{23.9}&\textbf{394.2}&\textbf{11}&\textbf{48}&\textbf{40}\\
\hline
6&20&36.7&0.0&0.0&3&10&10&19.0&0.0&0.0&3&10&10&8.1&0.0&0.0&4&10&10&21.3&0.0&0.0&10&30&30\\
6&40&65.2&18.1&19.2&0&2&1&45.0&13.5&16.1&0&2&1&230.8&61.8&150.5&0&2&1&113.7&31.1&61.9&0&6&3\\
6&60&102.4&11.5&625.0&0&4&0&141.0&15.2&657.5&0&0&0&399.3&218.9&3073.5&0&0&0&214.2&81.9&1396.1&0&4&0\\
6&80&61.6&1.7&2541.6&0&2&0&57.7&9.2&4408.7&0&0&0&409.3&46.8&17238.9&0&0&0&176.2&19.2&7746.7&0&2&0\\
6&100&69.0&15.4&6310.5&0&0&0&110.5&24.7&8378.1&0&0&0&192.7&57.2&16434.8&0&0&0&124.1&32.4&10374.4&0&0&0\\
\hline
\multicolumn{2}{l}{\textbf{Category 6}}&\textbf{67.0}&\textbf{9.3}&\textbf{1899.2}&\textbf{3}&\textbf{18}&\textbf{11}&\textbf{74.6}&\textbf{12.5}&\textbf{2692.1}&\textbf{3}&\textbf{12}&\textbf{11}&\textbf{248.0}&\textbf{76.9}&\textbf{7263.9}&\textbf{4}&\textbf{12}&\textbf{11}&\textbf{129.9}&\textbf{32.9}&\textbf{3907.0}&\textbf{10}&\textbf{42}&\textbf{33}\\
\hline
7&20&30.1&10.6&3.4&0&4&9&16.2&11.7&0.0&4&4&10&28.5&12.2&0.0&4&6&10&24.9&11.5&1.1&8&14&29\\
7&40&43.5&26.9&30.5&0&0&0&56.0&38.1&33.0&0&0&0&120.8&98.4&71.5&0&0&1&73.4&54.4&45.0&0&0&1\\
7&60&39.6&27.3&83.8&0&0&0&62.0&50.1&179.6&0&0&0&167.2&137.4&409.8&0&0&0&89.6&71.6&225.9&0&0&0\\
7&80&41.7&28.0&523.0&0&0&0&73.9&52.3&760.5&0&0&0&122.1&96.6&1837.6&0&0&0&79.2&59.0&1040.4&0&0&0\\
7&100&38.1&26.8&772.4&0&0&0&68.8&49.1&1324.1&0&0&0&235.4&209.6&4148.8&0&0&0&114.1&95.2&2081.8&0&0&0\\
\hline
\multicolumn{2}{l}{\textbf{Category 7}}&\textbf{38.6}&\textbf{23.9}&\textbf{282.6}&\textbf{0}&\textbf{4}&\textbf{9}&\textbf{55.4}&\textbf{40.3}&\textbf{465.1}&\textbf{4}&\textbf{4}&\textbf{10}&\textbf{134.8}&\textbf{110.8}&\textbf{1293.5}&\textbf{4}&\textbf{6}&\textbf{11}&\textbf{76.3}&\textbf{58.3}&\textbf{681.9}&\textbf{8}&\textbf{14}&\textbf{30}\\
\hline
8&20&16.5&3.8&0.1&0&4&9&12.8&3.2&0.0&3&6&10&67.3&55.7&0.0&0&2&10&32.2&20.9&0.0&3&12&29\\
8&40&43.4&28.6&30.5&0&0&0&59.3&42.0&38.4&0&0&0&136.3&96.3&83.3&0&0&1&79.6&55.6&50.7&0&0&1\\
8&60&45.4&24.9&97.6&0&0&0&60.5&41.8&130.7&0&0&0&423.6&294.0&824.1&0&0&0&176.5&120.2&359.5&0&0&0\\
8&80&40.9&25.6&566.6&0&0&0&72.7&54.4&847.4&0&0&0&399.2&307.5&4538.8&0&0&0&170.9&129.2&1984.3&0&0&0\\
8&100&38.6&25.4&684.0&0&0&0&72.6&51.9&1362.5&0&0&0&254.5&206.6&4570.2&0&0&0&121.9&94.6&2205.6&0&0&0\\
\hline
\multicolumn{2}{l}{\textbf{Category 8}}&\textbf{37.0}&\textbf{21.7}&\textbf{279.4}&\textbf{0}&\textbf{4}&\textbf{9}&\textbf{55.6}&\textbf{38.6}&\textbf{475.8}&\textbf{3}&\textbf{6}&\textbf{10}&\textbf{256.2}&\textbf{192.0}&\textbf{2003.3}&\textbf{0}&\textbf{2}&\textbf{11}&\textbf{116.2}&\textbf{84.1}&\textbf{923.8}&\textbf{3}&\textbf{12}&\textbf{30}\\
\hline
9&20&2.8&0.0&0.0&9&10&10&5.9&0.0&0.0&8&10&10&1.7&0.0&0.0&9&10&10&3.5&0.0&0.0&26&30&30\\
9&40&2.8&0.0&0.0&7&10&10&6.9&0.0&0.0&5&10&9&10.1&0.0&0.0&8&10&10&6.6&0.0&0.0&20&30&29\\
9&60&1.7&0.0&0.0&7&9&10&1.3&0.0&0.0&9&10&10&5.2&2.0&0.0&8&9&10&2.7&0.7&0.0&24&28&30\\
9&80&0.8&0.0&0.0&8&10&10&2.5&0.0&4.1&6&10&8&5.4&0.0&0.0&8&10&10&2.9&0.0&1.4&22&30&28\\
9&100&1.0&0.0&37.8&8&10&8&2.0&0.0&11.1&7&10&8&2.3&0.0&7.0&8&10&6&1.8&0.0&18.6&23&30&22\\
\hline
\multicolumn{2}{l}{\textbf{Category 9}}&\textbf{1.8}&\textbf{0.0}&\textbf{6.9}&\textbf{39}&\textbf{49}&\textbf{48}&\textbf{3.7}&\textbf{0.0}&\textbf{2.9}&\textbf{35}&\textbf{50}&\textbf{45}&\textbf{4.9}&\textbf{0.4}&\textbf{1.3}&\textbf{41}&\textbf{49}&\textbf{46}&\textbf{3.5}&\textbf{0.1}&\textbf{3.7}&\textbf{115}&\textbf{148}&\textbf{139}\\
\hline
10&20&25.4&13.4&12.5&3&6&7&37.9&26.4&16.2&2&6&8&229.0&165.4&161.1&1&4&7&97.4&68.4&63.3&6&16&22\\
10&40&29.9&14.2&22.1&0&1&0&50.2&33.2&39.8&0&1&1&209.0&153.6&172.3&1&4&4&96.4&67.0&78.0&1&6&5\\
10&60&32.4&17.0&214.4&0&0&0&52.0&36.5&166.6&0&0&0&86.3&66.1&645.2&0&0&0&56.9&39.8&330.4&0&0&0\\
10&80&31.3&18.7&397.9&0&0&0&54.3&35.4&1058.2&0&0&0&211.6&168.1&3409.9&0&0&0&99.0&74.1&1622.0&0&0&0\\
10&100&30.0&17.5&1295.2&0&0&0&47.2&32.3&1872.7&0&0&0&267.1&186.6&7995.1&0&0&0&114.8&78.8&3721.0&0&0&0\\
\hline
\multicolumn{2}{l}{\textbf{Category 10}}&\textbf{29.8}&\textbf{16.2}&\textbf{392.0}&\textbf{3}&\textbf{7}&\textbf{7}&\textbf{48.3}&\textbf{32.8}&\textbf{640.2}&\textbf{2}&\textbf{7}&\textbf{9}&\textbf{200.6}&\textbf{147.9}&\textbf{2553.0}&\textbf{2}&\textbf{8}&\textbf{11}&\textbf{92.9}&\textbf{65.6}&\textbf{1185.8}&\textbf{7}&\textbf{22}&\textbf{27}\\
\hline
\hline
\multicolumn{2}{l}{\textbf{All categories}}&\textbf{31.4}&\textbf{11.9}&\textbf{722.2}&\textbf{85}&\textbf{171}&\textbf{144}&\textbf{43.1}&\textbf{20.5}&\textbf{1042.8}&\textbf{89}&\textbf{164}&\textbf{146}&\textbf{126.1}&\textbf{68.5}&\textbf{2699.9}&\textbf{90}&\textbf{174}&\textbf{159}&\textbf{66.9}&\textbf{33.6}&\textbf{1483.7}&\textbf{264}&\textbf{509}&\textbf{449}\\
\hline
\end{tabular}
}
\end{tiny}
\end{sidewaystable}

Table \ref{Tab:UB} infers the following results. The mean $L_{max}^{\bullet},\ \bullet=F,\ A,\ E,$ is equal for both classes A and B and larger for class C.
The average $\delta^{F}$ and the average $\delta^{A}$ are larger than their respective medians (i.e., 66.85\% and 33.62\% versus 37.91\% and 9.89\%, respectively); signaling few outliers that are enlarging the true size of $\delta^{F}$ and $\delta^{A}$. This is expected from $\mathcal{NP}$-hard problems.

The mean $\delta^{A}$ is the smallest. Its point and confidence interval estimates are 33.62\% and (28.57\%, 38.67\%).  That is, on average, the application of the second phase of the algorithm improves the solution of {\FG} (except when $L_{max}^{F}=L_{max}^{A}=L_{max}^{*}$). The mean improvement is of the order of 33.23\%, with a 29.59\% lower side estimate.  On the other hand, the mean $\delta^{E}$ is the largest because {\MIP} fails to obtain reasonably good solutions for large instances.

There is no correlation between $\delta^{F},\ \delta^{A},\ \eta^{F},\ \eta^{A}$ and $n$, but there is a moderate correlation between both $\delta^{E},\ \eta^{E}$ and $n$ with respective 0.402 and -0.598 Pearson correlation coefficients. This infers that as the problem size increases, {\MIP} may encounter increasing difficulty in getting the tightest upper bound.  $\delta^{E}$ depends on the problem category. Its mean for categories 2, 4, and 6 are larger than those for the other categories. Similarly, $\eta^{E}$ is category dependent. Its estimate is largest for category 9 and smallest for category 2. Finally, $\delta^{A}$ is category dependent. Its mean $\delta^{A}$ is smallest for categories 2 and 9 and largest for categories 7, 8, and 10. {\Alga} solves many of the instances of categories 2 and 9 to optimality. It neither reaches the optimum nor proves the optimality of its solutions for any of the instances of categories 7 and 8. Even though category dependent, the mean number of times optimality is proven does not differ among sets or classes.

A valid upper bound to $L_{max}$ is the minimum of $L_{max}^A$ and $L_{max}^E.$ This upper bound equals $LB^*$ for 586 out of 1500 instances; that is, in 39.07\% of the cases.

%>>>>>>>>>>>>>>>>>>>>>>>>>>>>>>>>>>>>>>>>>>>>>>>>>>>>>>>>>>>>>>>>>>>>>>>>>>>>>>>>>>>>>>>>>>>>>>>>>>>>>>>>>>>>>>>>>
\subsection{Comparing {\FG} and {\Alga} to {\GA}}\label{ssec:CEGA}

This section investigates the performance of {\Alga} relative to existing upper bounds. Table \ref{tab:res} reports the results per class and category as \citet{Bennell13} do not provide results per problem size. Column 3 gives the average number of attempts made by {\APr} (and {\APrr}) in order to reach a feasible solution within a single run. Columns 4-6 and 7-9 report statistics of the runtime $RT^{\bullet},\ \bullet=F,\ A,$ in seconds: the average, median, and maximum run time, all in seconds, over each set of 50 instances. $RT^A$ includes $RT^F$ as it pre-calls {\FG}. $RT^{\GA}$ is fixed to 120 seconds per replication for each of the ten replications; thus, is not included in the table. Columns 10-11 give $\#^{\bullet},\ \bullet=F,\ A,$ the number of times the run times of {\FG} and {\APr} are larger than the 120 second runtime of one replication of {\GA}.  This number is out of 50 for each class and category and out of 500 for each class.  Finally, Columns 12-14 report the average relative percent gap $\gamma^{\bullet}=100 {(L_{max}^{\bullet}-\BSP)}/{\BSP},\ \bullet=F,\ A,\ {\GA}.$

\begin{table}[htb]
\centering
\caption{Summary results of computational experiments}\label{tab:res}
{\tiny
\begin{tabular}{rr|r|rrr|rrr|rr|rrr}
Class&Category&iter&\multicolumn{3}{c|}{$RT^F$}&\multicolumn{3}{c|}{$RT^A$}&
$\#^F$&$\#^{A}$&$\gamma^F$&$\gamma^A$&$\gamma^{\GA}$\\
& & &Mean&Median&Max&Mean&Median&Max& & & & &\\
\hline
A&1&25&0.0&0.0&0.6&18.4&13.0&69.4&0&0&28.4&13.9&\textbf{12.4}\\
&2&12&11.5&8.8&35.6&20.7&14.3&69.0&0&0&12.8&\textbf{2.3}&11.1\\
&3&24&2.6&1.7&11.6&25.1&18.6&78.0&0&0&46.7&22.1&\textbf{22.0}\\
&4&14&34.0&29.1&87.6&60.4&52.5&197.0&0&8&36.8&\textbf{10.5}&17.1\\
&5&29&2.2&1.7&12.0&26.0&17.9&126.2&0&1&37.9&19.1&\textbf{17.9}\\
&6&24&46.1&42.7&136.2&76.2&58.7&228.8&2&14&67.0&\textbf{9.3}&16.6\\
&7&33&0.8&0.4&7.0&24.6&20.5&82.6&0&0&43.0&27.7&\textbf{23.5}\\
&8&40&1.1&0.4&7.4&27.3&18.3&105.0&0&0&42.2&26.5&\textbf{23.3}\\
&9&2&0.1&0.0&2.0&12.7&0.1&79.8&0&0&3.5&\textbf{1.7}&1.7\\
&10&19&13.3&7.9&76.2&26.0&17.8&108.4&0&0&34.3&\textbf{20.2}&23.8\\
\hline
\multicolumn{2}{l|}{All}&&	11.2&1.2&136.2&31.7&19.6&228.8&2&23\\
\hline
B&1&23&0.0&0.0&0.0&15.7&10.0&73.6&0&0&47.5&\textbf{23.1}&24.2\\
&2&3&10.3&7.6&50.0&19.1&13.5&74.4&0&0&9.0&\textbf{2.9}&34.0\\
&3&20&2.5&1.4&12.2&17.5&11.7&78.0&0&0&77.8&\textbf{41.7}&46.2\\
&4&21&28.1&20.4&82.0&50.8&35.4&168.8&0&7&40.9&\textbf{14.3}&36.0\\
&5&32&2.2&1.0&14.6&17.8&9.2&154.4&0&1&61.5&\textbf{33.3}&35.5\\
&6&22&39.1&35.4&134.4&63.4&53.3&292.4&2&10&74.6&\textbf{12.5}&37.7\\
&7&46&0.8&0.2&4.8&21.0&9.8&81.8&0&0&64.2&\textbf{48.7}&52.2\\
&8&47&0.7&0.4&3.8&21.1&10.9&107.0&0&0&64.2&\textbf{46.5}&49.4\\
&9&2&0.1&0.0&1.2&8.1&0.0&61.8&0&0&6.2&\textbf{2.4}&2.4\\
&10&21&11.1&6.7&45.6&21.0&14.8&66.4&0&0&57.4&\textbf{40.9}&53.5\\
\hline
\multicolumn{2}{l|}{All}&&	9.5&0.8&134.4&25.5&11.5&292.4&2&18\\
\hline
C&1&24&0.0&0.0&0.4&11.5&7.9&44.0&0&0&117.9&\textbf{69.6}&93.0\\
&2&14&10.9&9.7&40.2&18.9&15.1&62.0&0&0&23.1&\textbf{7.2}&149.5\\
&3&24&2.3&1.7&7.2&20.2&13.5&75.8&0&0&152.0&\textbf{91.2}&124.9\\
&4&20&33.1&25.5&88.8&57.6&50.4&217.0&0&8&140.0&\textbf{35.6}&153.2\\
&5&28&3.0&2.0&12.0&23.5&12.8&85.0&0&0&146.7&\textbf{81.1}&105.0\\
&6&29&44.6&45.3&139.0&77.8&68.1&278.4&2&12&248.0&\textbf{76.9}&241.2\\
&7&41&1.2&0.6&7.4&25.3&12.8&108.8&0&0&186.3&\textbf{156.7}&209.6\\
&8&37&1.8&0.8&11.0&26.7&17.1&149.4&0&1&301.5&\textbf{232.2}&273.3\\
&9&2&0.2&0.0&1.8&10.4&0.2&66.8&0&0&15.7&10.5&\textbf{9.9}\\
&10&18&15.0&11.1&47.8&30.1&21.2&156.6&0&1&282.9&\textbf{214.5}&318.5\\
\hline
\multicolumn{2}{l|}{All}&&	11.2&1.6&139.0&30.2&16.1&278.4&2&22\\
\end{tabular}
}
\end{table}

Table \ref{tab:res} shows that {\APr} needs on average 23 attempts to reach a feasible solution. The mean number of attempts differs among categories. It is smaller for categories 2 and 9 with respective point estimates of 9.67 and 2.00. The instances of these two categories contain many tiny items. Therefore, packing these items is relatively easy. The mean number of attempts is larger for categories 7 and 8 with respective point estimates of 40.00 and 41.33. However, it is not different among sets $\mathcal{S}$ and $\mathcal{L}$; that is, what defines the level of difficulty of packing is the homogeneity of the items rather than their sizes. Finally, this mean number of attempts does not differ among classes. This is expected because the differences of classes is caused by the tightness of the due dates and not by the packing procedure itself.

\begin{figure}[htb]
\centering
\includegraphics[width=\textwidth,height=7.0cm]{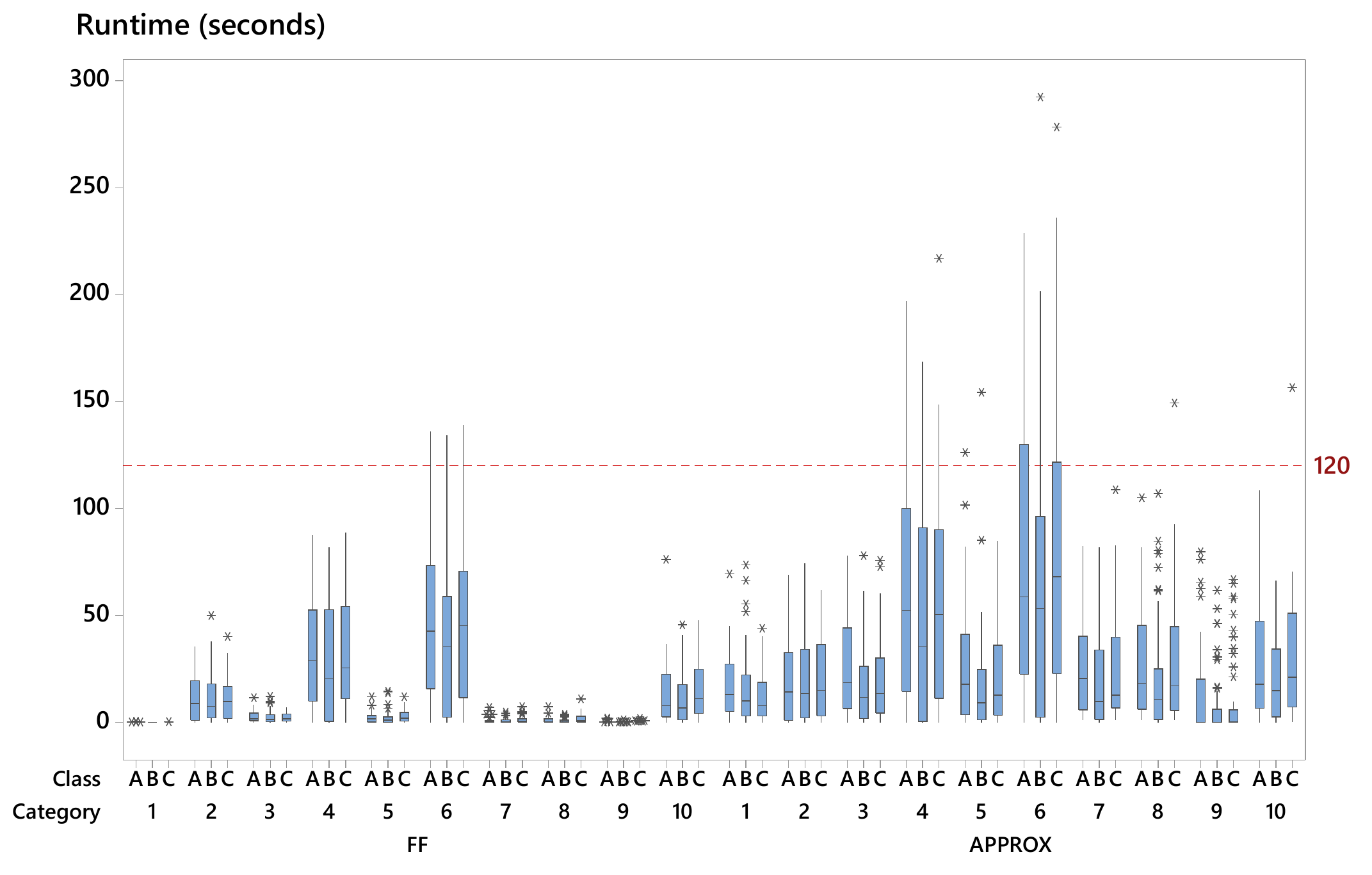}
 \caption{Box plot for the run times of {\FG} and {\Alga} by category and class}\label{Fig:RT}
\end{figure}

\begin{figure}[htb]
\centering
\includegraphics[width=\textwidth]{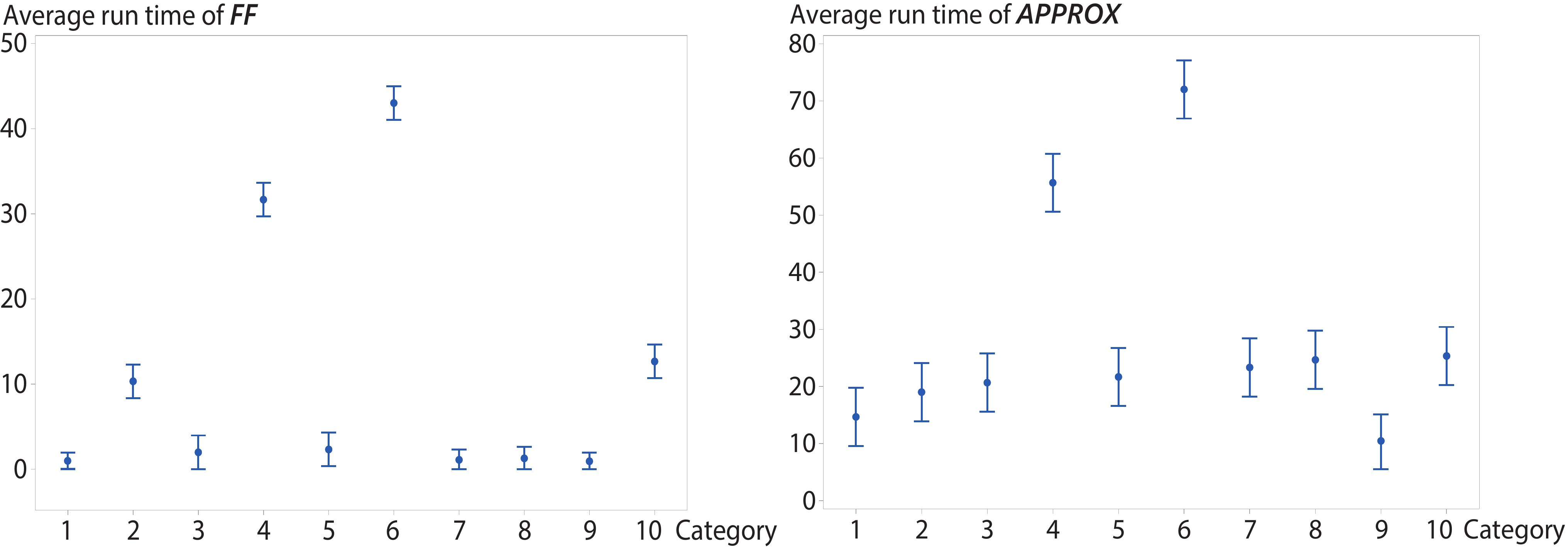}
 \caption{95\% confidence intervals for the mean run times of {\FG} and {\Alga} by category}\label{figRT}
\end{figure}

{\FG} and {\Alga} identify a local optimum in 10.6 and 29.2 seconds, on average.  These values are inflated by few outliers, as the box plots of Figure \ref{Fig:RT} illustrate.  In fact, the respective median run times are 1.2 and 15.6 seconds.
Over all 1500 tested instances, {\FG} needed a larger than 120-second runtime for 6 instances.  Similarly, {\Alga} needed a larger than 120-second runtime for merely 63 out of 1500 instances; that is, in 4.2\% of the tested cases. Figure \ref{figRT} displays the confidence interval estimates of the mean run times per category of {\FG} and {\Alga}. The mean run times of both {\FG} and {\Alga} are less than the 120-second runtime of {\GA} at any level of significance. That is, {\Alga} is, on average, faster than {\GA}.

{\FG} solves the instances of set $\mathcal{L}$ particularly fast. However, it is relatively slower on the instances of set $\mathcal{S}$. In fact, the large number of items per bin for instances of $\mathcal{S}$ leads to inaccurate lower bounds {\DMV}; thus, increases the number of calls to {\CPr}, which checks the feasibility of a packing. In addition, this large number of items slows down {\CPr}. This, in turn, translates into increased run times. {\Alga} spends, on average, approximately the same time to solve an instance of any category, except of categories 4 and 6.

{\FG} is as competitive as {\GA} in terms of solution quality. Their mean optimality gaps are not different.  However, {\FG} is faster than {\GA}. In addition, {\FG} outperforms {\GA} on some categories of classes B and C.

\begin{figure}[htb]
\begin{center}
\begin{tabular}{cc}
\adjustimage{height=5cm}{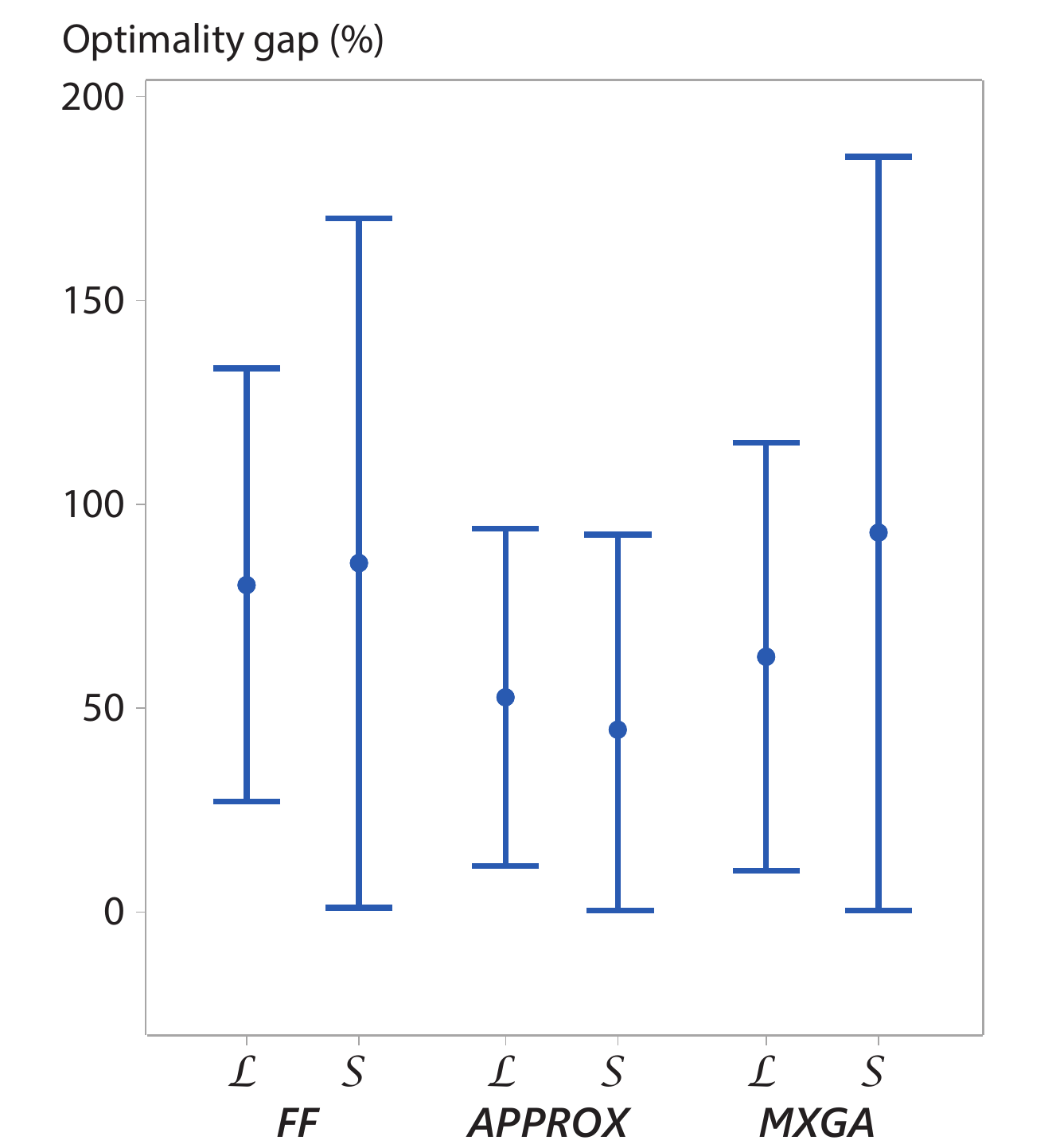} & \adjustimage{height=5cm}{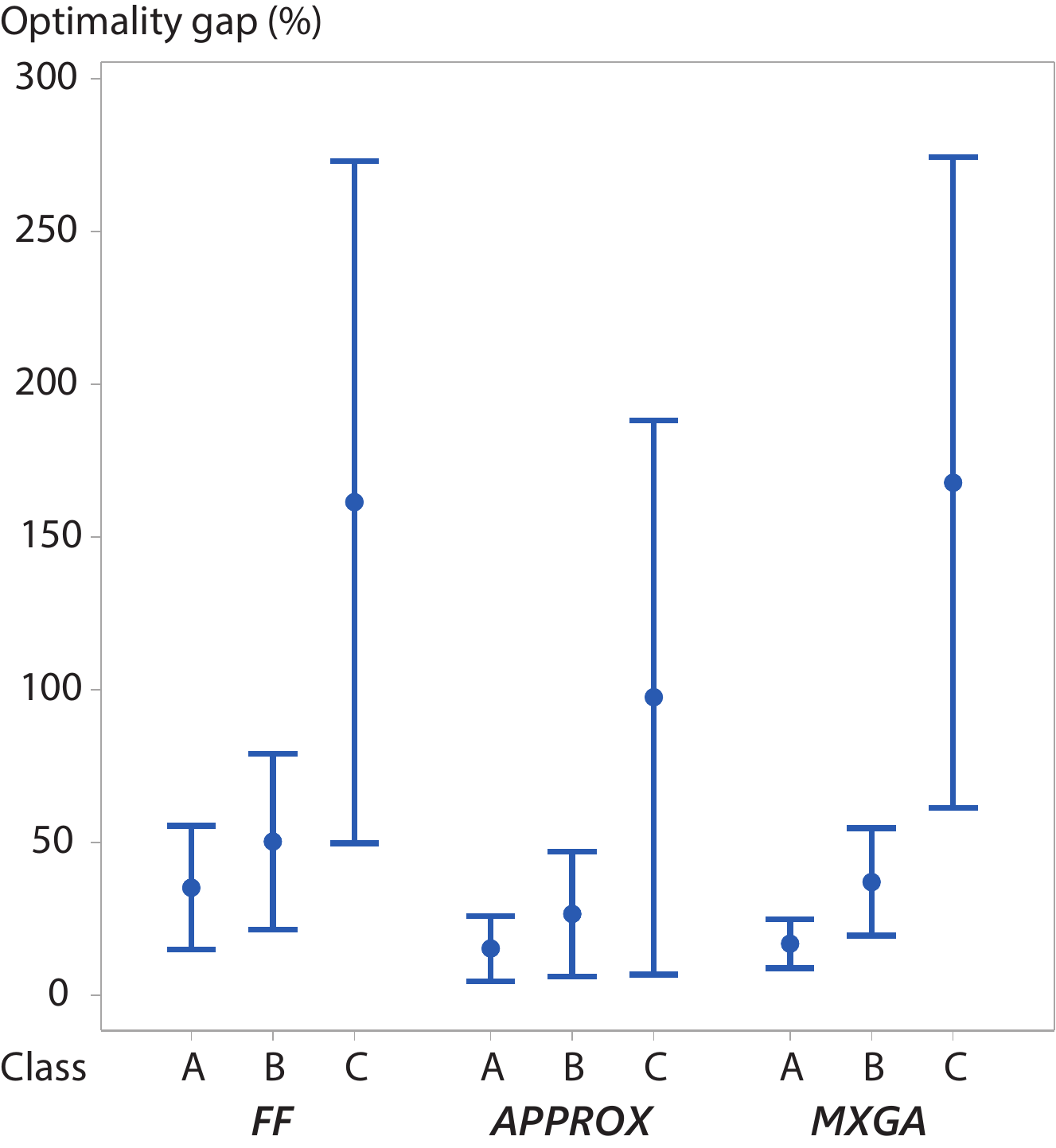}\\
a)&b)
\end{tabular}
\caption{95\% confidence interval of the mean optimality gaps obtained by {\FG}, {\Alga} and {\GA} (a) by set (b) and by class.}
\label{Fig3}
\end{center}
\end{figure}

\begin{figure}[htb]
 \centering
 \includegraphics[width=8cm]{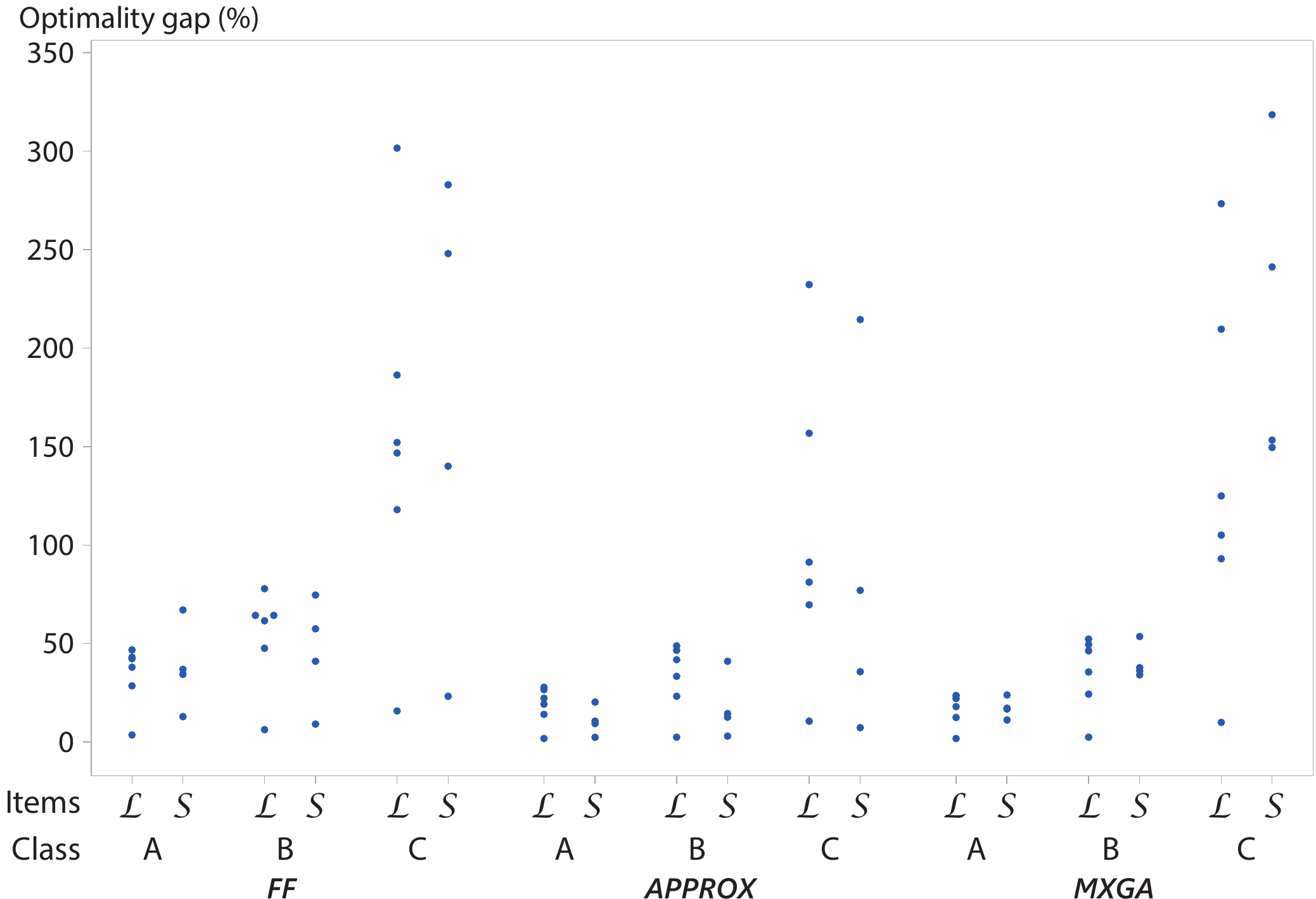}
 \caption{Observed percent optimality gaps of {\FG}, {\Alga} and {\GA} by class}\label{FigLSclass}
\end{figure}

{\Alga} outperforms {\GA} on average.  The mean optimality gap of {\Alga} is smaller than its {\GA} counterpart, with point and confidence interval estimates of the mean difference of -27.45\% and (-44.22\%,-10.68\%). As further substantiated by Figure \ref{Fig3}, the mean optimality gap of {\Alga} is smaller than its {\GA} counterpart, for both sets $\mathcal{L}$ and $\mathcal{S}$ and for all classes. Indeed, the mean difference depends on the class with -1.61, -10.48, and -70.30\% point estimates for classes A, B, and C, respectively.  These results are most likely due to the nature of {\GA}. In fact, {\GA} along with its packing algorithm are strongly oriented towards obtaining a dense packing; a rather important criterion for class A whose instances are characterized by narrow intervals of due dates. However, {\GA} is myopic when the due dates are sparse. This myopic nature is further highlighted by Figure \ref{FigLSclass}, which shows that {\Alga} produces better results than {\GA} more frequently, in particular for class C whose items have a wider range of due dates.

The size of the optimality gaps seems unusually high. Even though part of it may be due to the quality of the upper bounds obtained by {\FG}, {\Alga} and {\GA}, most of it is most likely due to the looseness of the lower bound. The comparison of Tables \ref{Tab:UB} and \ref{tab:res} further supports this claim. Tables \ref{Tab:UB} and Table \ref{tab:res} report the percent deviations of {\FG} and {\Alga} from the best lower bound $LB^*$ and {\BSP}, respectively. For {\FG}, this gap is 282.9\% for class C, category 10 in Table \ref{tab:res}, but reduces to 204\% in Table \ref{Tab:UB}. A good example is rand10.txt\_C\_1, the 10th instance of category 1, class C, $n=20$. The percent gaps of {\FG} and {\Alga} are 136\% when computed with respect to {\BSP}$= 69$, but become 0\% when computed with respect to {\PH}$= 163.$ In this case, the upper bound matches {\PH}, and proves the optimality of the solutions obtained by {\FG} and {\Alga}. Even though the instances are numerous, we only cite a second example: rand6.txt\_C\_10 of category 10, class C, $n=20,$ where {\BSP}$= 3$ results in an optimality gap of 1533\% whereas {\PH}$=92$ proves the optimality of the upper bound obtained by {\Alga}.

In summary, the proposed approach enhances many existing upper bounds, assesses the tightness of existing and proposed lower bounds, and proves the optimality of many open benchmark problems. It outperforms {\GA} in terms of solution quality, run time, and number of proven optima.

%>>>>>>>>>>>>>>>>>>>>>>>>>>>>>>>>>>>>>>>>>>>>>>>>>>>>>>>>>>>>>>>>>>>>>>>>>>>>>>>>>>>>>>>>>>>>>>>>>>>>>>>>>>>>>>>>>
\subsection{Performance of {\FG} and {\Alga} on Large-Sized Instances}\label{ssec:scalability}

This section investigates the performance of {\FG} and {\Alga} for very large instances. For this purpose, it expands the current benchmark set of $n=100$ to a set with $n= 100 \tau,\ \tau=2,\ldots,5,$ by duplicating the items of each instance $\tau$ times. It creates 10 new instances per class, category and problem size; generating the due dates for the new items as Section~\ref{ssec:CEsetup} explains.

For large-sized instances, {\FG} may require a sizeable number of iterations to reach a feasible solution. A large ratio of number of items per bin makes the lower bound in {\FG} loose; thus ineffective for the packing test. As a result, {\CPr} is called numerous times with most of these calls being useless due to their large problem size. {\FG} checks feasibility after appending a single item. Every feasibility check increases the runtime. A number of heuristic strategies can be designed to limit such ineffective calls. \textbf{Two} such strategies follow.

The \textbf{simplest} strategy limits the search of Loop~(\ref{code:1}-\ref{code:2}) of Algorithm~\ref{alg:1} to $\sigma$ items among the items that succeed item $i$ in $N$. This involves a new counter, which is incremented every time Condition (\ref{code:3}) fails, and is re-initiated to zero otherwise. The loop stops when the counter equals $\sigma$.

A \textbf{second} strategy monitors the maximal dimension $\mu$ of an item that can be packed in the current bin $k$. Initially, $\mu=max\left\{W,H\right\}$. When item $i$ violates Condition (\ref{code:3}), a supplementary test ${\CPr}(N_k \cup \left\{i'\right\})$ is run to check the feasibility of packing a dummy item $i'$ defined by $w_{i'}=max\left\{w_i,h_i\right\}$ and $h_{i'}=1$ into $k$. If the test reveals the infeasibility of such a packing, $\mu$ is updated to $w_{i'}$, and Loop~(\ref{code:1}-\ref{code:2}) skips any subsequent item $i$ whose $max\left\{w_i,h_i\right\}\geq\mu$. The additional calls to {\CPr} increase the runtime of the heuristic; a negligible increase compared to the runtime saved by the omitted unsuccessful future calls to {\CPr}. That is, this strategy is efficient only when $n$ is large. Because {\CPr} is a heuristic, its outcome may suggest an infeasible packing while a feasible one exists. In fact, the choice of $\mu$ is not based on an exact method but on a rule of thumb. Thus, the application of {\CPr} may slightly degrade the solution quality.

Similarly, {\Alga} may require too many iterations of {\APr} (and {\APrr}), with each resulting in a slight improvement of {\U}. To avoid a series of too many small enhancements, {\Alga} adopts the minimal improvement technique, which first computes a value for {\U} using {\FG}, and iteratively examines feasible solutions whose lateness is $\delta \%$ smaller than the current {\U}. When it fails to find a solution whose lateness is $\delta \%$ better than the incumbent, it examines solutions sequentially (as Section~\ref{sec:SC} explains) until no further improvement can be achieved.

\begin{figure}[htb]
\begin{center}
\begin{tabular}{ccc}
\adjustimage{width=0.3\textwidth}{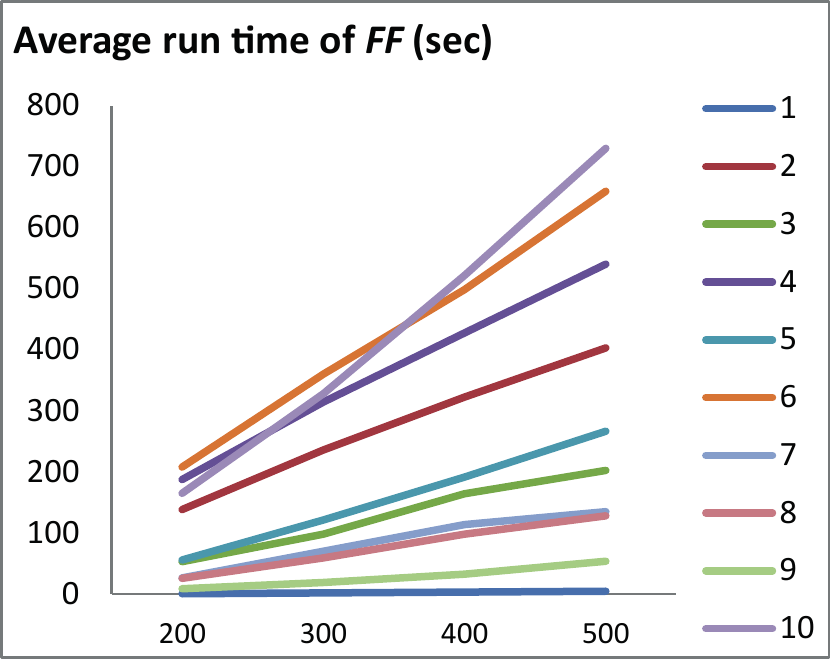} & \adjustimage{width=0.3\textwidth}{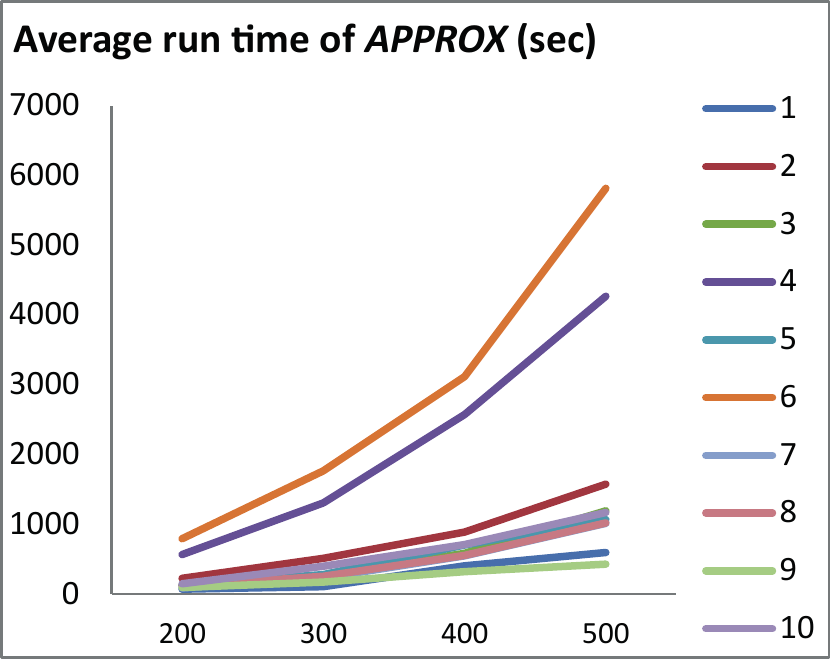} & \adjustimage{width=0.3\textwidth}{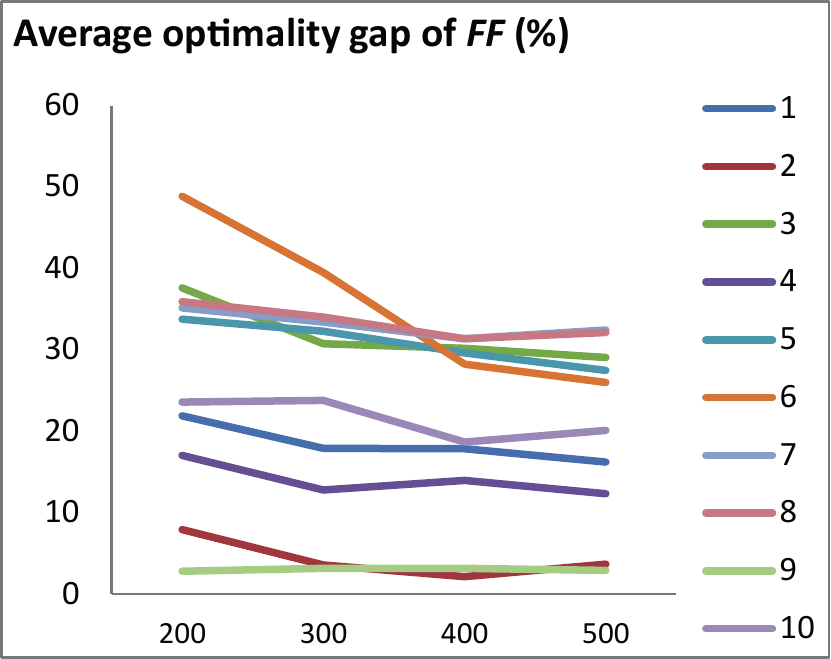} \\
\small{a})&\small{b})&\small{c})\\
\adjustimage{width=0.3\textwidth}{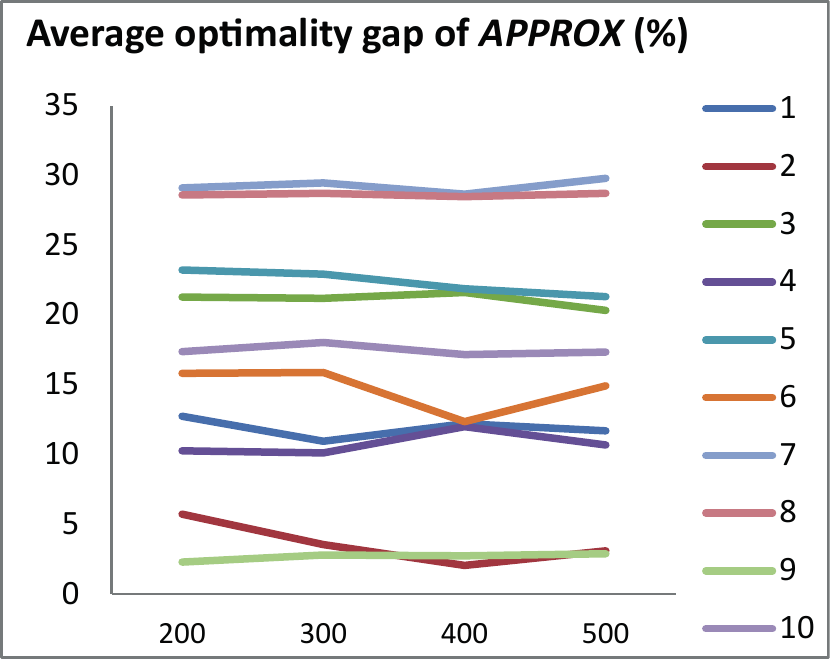} & \adjustimage{width=0.3\textwidth}{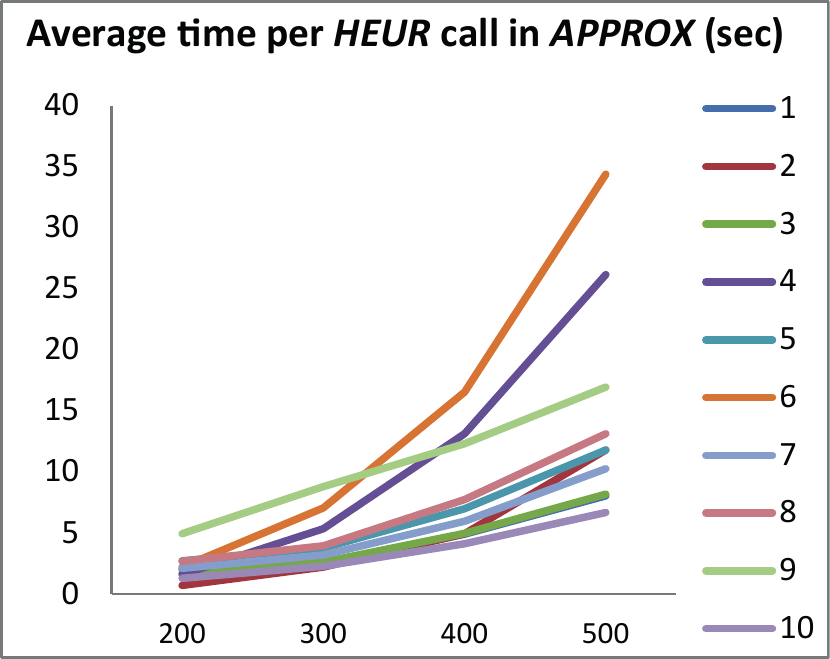} & \adjustimage{width=0.3\textwidth}{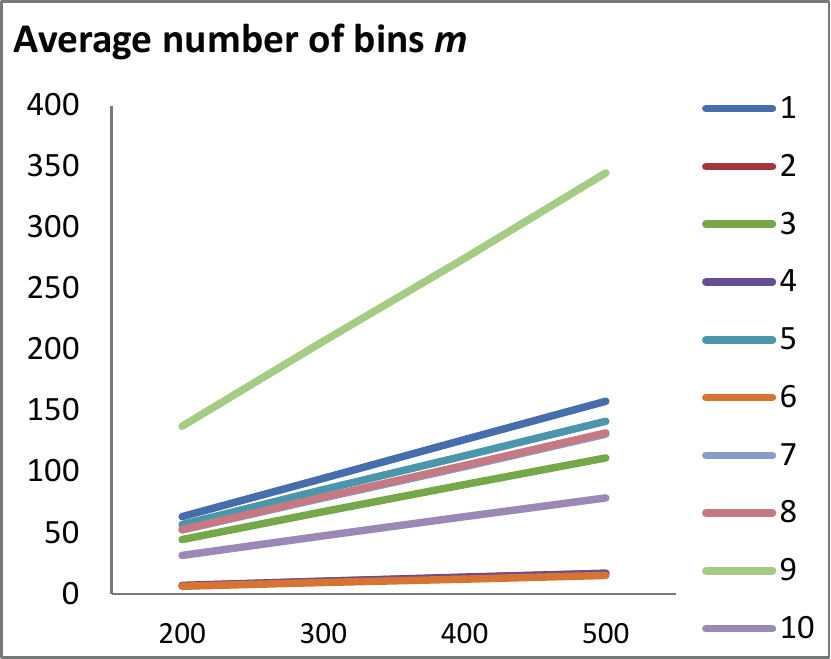} \\
\small{d})&\small{e})&\small{f})\\
\end{tabular}
\caption{Summary results for large-sized instances of class A}
\label{tb:scalability}
\end{center}
\end{figure}

Herein, {\FG} and {\Alga} are run with a 3-second time limit for {\CPr}, and maximal {\co} number of iterations \text{\LHEUR}=30 and  $a_{lim}^{\text{\textsl{HEUR'}}}=10$. In addition, the two heuristic strategies are applied for instances of set $\mathcal{S}$ with $\sigma=40$. Despite their slightly weaker solutions' quality, the two strategies reduce the runtime by an order of magnitude for instances with 500 items. Finally, the minimal improvement technique is used with $\delta=2\%.$  It speeds significantly the search for a local optimum for instances with $n>100$. The summary of the results for class A is displayed in Figure~\ref{tb:scalability}.  The results for the other two classes behave similarly, and are therefore omitted from further consideration.  Figures~\ref{tb:scalability}.a-d depict the average run times for {\FG} and {\Alga} along with their optimality gaps, computed with respect to {\BSP}. Figure~\ref{tb:scalability}.e illustrates the average runtime per {\APr} call in {\Alga}. This runtime sums the times required by the iterative calls to {\ASSIGN} until either a feasible solution is reached or infeasibility is detected. Finally, Figure~\ref{tb:scalability}.f reports the average number of bins used per problem size and category.

Figure~\ref{tb:scalability}.a shows the linear growth of the average runtime of {\FG} as a function of the number of items with a near to one correlation of determination. This trend is in turn dictated by the perfectly linear growth of the number of bins as $n$ increases, as Figure~\ref{tb:scalability}.f illustrates. This is expected as the time limit for {\CPr} is a prefixed constant.  To the opposite and as Figure~\ref{tb:scalability}.b suggests, the average runtime of {\Alga} is not linearly proportional to $n$ as there is no time limit on {\MIPG}.  Categories 4 and 6 have the hardest instances. Their mean run times differ from those of all other categories.  These two categories have the largest number of items per bin ratio as well as the largest variable domains that are used by {\CP} and {\MIPG}. Subsequently, they both require a larger mean runtime per {\APr} call as Figure~\ref{tb:scalability}.e shows, and a larger number of {\APr} calls while the {\ASSIGN} sub-problem itself is harder. The mean run times of both {\FG} and {\APr} are category, set and size dependent.

The mean optimality gap of {\FG} is category dependent. It is not however set or size dependent. On the other hand, the mean optimality gap of {\APr} is both category and set dependent but is not size dependent. {\APr} reduces the optimality gap of all the solutions obtained by {\FG}. This can be further inferred by the comparison of Figures~\ref{tb:scalability}.c and \ref{tb:scalability}.d. The mean reduction is 4.61\%. It is neither set nor size dependent but is category dependent. It is largest for category 6 with point and 95\% confidence interval estimates of 9.61\% and (17.21\%,24.64\%), respectively.

%%%%%%%%%%%%%%%%%%%%%%%%%%%%%%%%%%%%%%%%%%%%%%%%%%%%%%%%%%%%%%%%%%%%%%%%%%%%%%%%%%%%%%%%%%%%%%%%%%%%%%%%%%%%%%%%%%
\section{Conclusion}\label{sec:Conclusion}
%%%%%%%%%%%%%%%%%%%%%%%%%%%%%%%%%%%%%%%%%%%%%%%%%%%%%%%%%%%%%%%%%%%%%%%%%%%%%%%%%%%%%%%%%%%%%%%%%%%%%%%%%%%%%%%%%%

This paper addresses the two-dimensional non-oriented bin packing problem with due dates. It proposes a lower bound, an exact mixed integer model, and an approximate approach that significantly enhances existing results on many benchmark instances from the literature. It solves 33.93\% of the instances to optimality.  Because the exact model can be solved to optimality by an off-the-shelf solver, the total percent of instances solved to optimality is 39.07\%.  Unlike many traditional constructive packing heuristics, the packing approach packs simultaneously several items into several bins and takes advantage of the feasibility constraints to guide the search to a local optimum. Its concept of free regions is not specific to the due dates complicating constraints. It makes the proposed approach easily adaptable to other complex bin-packing related problems with problem-specific constraints such as routing, time windows, and tardiness related costs. Since the dual feasible functions are applicable to higher-dimensional packing, the approach can also be extended to this area.

\bibliographystyle{model5-names}
\bibliography{references}

\end{document}